\documentstyle[amssymb,aps, manuscript, graphics]{revtex}

\begin{document}
\title{Mesoscopic fluctuations of off-diagonal matrix elements of the angular
momentum and orbital magnetism of free electrons in a rectangular box.}
\author{M. X. Lou, J. M. A. S. P. Wickramasinghe and R. A. Serota}
\address{Department of Physics\\
University of Cincinnati\\
Cincinnati, OH\ 45221-0011}
\maketitle

\begin{abstract}
We study, analytically and numerically, mesoscopic fluctuations of the
off-diagonal matrix elements of the orbital angular momentum between the
nearest energy levels $i=\left( n_{x},n_{y}\right) $ and $f=\left(
k_{x},k_{y}\right) $ in a rectangular box with incommensurate sides. In the
semiclassical regime, where the level number of $i$ is ${\cal N}\gg 1$, our
derivation gives $\left\langle \left| \widehat{L}_{if}\right|
^{2}\right\rangle \sim \sqrt{{\cal N}}$. Numerical simulations, using
simultaneous ensemble averaging (over the aspect ratios of rectangles) and
spectral averaging (over the energy interval), are in excellent agreement
with this analytical prediction. Physically, the mean is dominated by the
level pairs $k_{x}=n_{x}\pm 1$, $k_{y}=n_{y}\mp 1$. Also in a rectangular
box, we investigate the mean orbital susceptibility of a free electron gas
and argue that it reduces, up to a coefficient, to the two-level van Vleck
susceptibility that involves the last occupied (Fermi) level $i$ and the
first unoccupied level $f$. This result is confirmed numerically as well,
albeit the effect of fluctuations is much more pronounced for the
susceptibility since it is due both to large fluctuations in $\left\langle
\left| \widehat{L}_{if}\right| ^{2}\right\rangle $ and in level separations $%
\varepsilon _{f}-\varepsilon _{i}$ (level bunching).
\end{abstract}

\section{Off-diagonal matrix elements of the angular momentum}

\subsection{Introduction}

Off-diagonal matrix elements of the orbital angular momentum enter into
important physical quantities, such as magnetic dipole absorption and van
Vleck susceptibility. This is particularly significant in the situations
when the angular momentum is not a good quantum number. Such is the case in
disordered systems where, in the semiclassical approximation, it was shown
that in 2D 
\begin{equation}
\left\langle \left| \widehat{L}_{if}\right| ^{2}\right\rangle \sim
\varepsilon _{F}\tau \sim k_{F}\ell  \label{Lif_dirt}
\end{equation}
Here $k_{F}$ is the wave number of a free particle whose energy $\varepsilon
_{F}$ corresponds to level $i$, and $\ell $ and $\tau $ are, respectively,
the mean-free-path and the scattering time due to disorder. This result can
be derived either by considering the classical magneto-dipole absorption\cite
{S1},\cite{MGL} or by a direct evaluation\cite{S1} using the technique
developed in Refs.\cite{Sh},\cite{GE}.

Disordered systems are classically chaotic. In the semiclassical regime they
exhibit ''level rigidity,'' which prevents large fluctuations in the level
spacings and, in turn, large fluctuations of the number of levels in an
energy interval.\cite{G}-\cite{E} Classically integrable systems, on the
other hand, exhibit ''level bunching'' characterized by large fluctuations
in level spacings\footnote{%
In the absence of resonances, or supersymmetry, such as the case for
harmonic oscillator and the Kepler problem.} and high occurrence of small
spacings - hence the term -\ and, in turn, large fluctuations in the number
of levels in an energy interval.\cite{G},\cite{BFFMPW} (It was recently
shown that such behavior extends only up to a certain energy scale, upon
which strong correlations between levels set in\cite{WGS}; Even then, the
number level variance exhibit large, non-decaying oscillations around the
''saturation value.'') Consequently, the mesoscopic, or non self-averaging,
effects are expected to be more pronounced in integrable systems.

The fluctuations in the level spacings (and more generally, the specifics of
correlations between the levels) is, however, just one of the factors
contributing to the mesoscopic fluctuations of physical quantities. Other
contributors are expected to fluctuate much stronger in classically
integrable systems than in classically chaotic systems as well. In the first
part of this paper we consider the fluctuations of the off-diagonal matrix
elements of the angular momentum in a rectangular box. We will show, in
particular, that 
\begin{equation}
\left\langle \left| \widehat{L}_{if}\right| ^{2}\right\rangle \sim \sqrt{%
{\cal N}}\sim k_{F}L  \label{Lif_rect}
\end{equation}
where ${\cal N}\gg 1$ is the level number of $i$ and $L$ is a rectangle's
side.

Brackets in eq. (\ref{Lif_dirt}) denote averaging over various realizations
of disorder, which is an example of ''ensemble averaging.'' A natural
extension of the concept of ensemble averaging to a rectangular box would be
to average over the aspect ratios of the rectangles' sides. However, our
analytical derivation of eq. (\ref{Lif_rect}) is based on ''spectral
averaging,'' that is averaging over an energy interval. A detailed
explanation of our numerical procedure will be given in text, but it should
be already mentioned that a {\it combined }ensemble and spectral averaging
was performed. The former involves averaging over the aspect ratios chosen
to be algebraic and close to $1.$ The latter is over an energy interval that
includes a large number of pairs $\left( i,f\right) $ and is, in fact,
necessary due to large fluctuations in $\left| \widehat{L}_{if}\right| ^{2}$%
. In other words, spectral averaging proves essential both in order to
derive a closed form analytical expression and to ensure convergence of
numerical results. Clearly, an underlying assumption is the validity of the
''ergodic hypothesis'' - that the two averages are equivalent.

This Section is organized as follows. First, we show that the magnitude of $%
\left| \widehat{L}_{if}\right| ^{2}$ is determined by the hierarchy of (odd)
pairs $\left( k_{x}-n_{x}\right) \ $and $\left( k_{y}-n_{y}\right) $, where $%
i=\left( n_{x},n_{y}\right) $ and $f=\left( k_{x},k_{y}\right) $ are the
nearest energy levels. For a given pair, $\left| \widehat{L}_{if}\right|
^{2}\propto {\cal N}$ with the largest coefficient, by two orders of
magnitude, corresponding to $k_{x}=n_{x}\pm 1$, $k_{y}=n_{y}\mp 1$.
Moreover, we show that the probability of the latter $\propto {\cal N}
^{-1/2} $, while for most pairs it is $\propto {\cal N}^{-1}$. Consequently,
such pairs give an overwhelming contribution to the spectral average, which
results in eq. (\ref{Lif_rect}). This is subsequently verified numerically.

\subsection{Spectrum and $\left| \widehat{L}_{if}\right| ^{2}$ in a
rectangular box}

For a rectangle, the energy eigenvalues are 
\begin{equation}
\varepsilon _{n_{x}n_{y}}=\frac{\pi ^{2}\hbar ^{2}}{2m}\left( \frac{n_{x}^{2}%
}{L_{x}^{2}}+\frac{n_{y}^{2}}{L_{y}^{2}}\right)  \label{enn}
\end{equation}
We consider rectangles of the same area $A=L_{x}L_{y}$ in an ensemble with
different values of the ratio $\alpha =L_{x}^{2}/L_{y}^{2}$. Numerically, we
use algebraic numbers for $\alpha $ to reduce accidental level degeneracies.
Expressing the energies in terms of the mean level spacing, 
\begin{equation}
\Delta =\frac{2\pi \hbar ^{2}}{mA}  \label{Delta}
\end{equation}
we find the dimensionless form of the spectrum 
\begin{equation}
\varepsilon _{n_{x}n_{y}}=\frac{\pi }{4}\left( \alpha
^{-1/2}n_{x}^{2}+\alpha ^{1/2}n_{y}^{2}\right) =\frac{\pi }{4}\alpha
^{-1/2}\left( n_{x}^{2}+\alpha n_{y}^{2}\right)  \label{enn_reduced}
\end{equation}
For $\alpha \sim 1$, we have a simplified expression for the spectrum 
\begin{equation}
\varepsilon _{n_{x}n_{y}}\approx \frac{\pi }{4}\left(
n_{x}^{2}+n_{y}^{2}\right) \equiv \frac{\pi }{4}N  \label{enn_reduced-1}
\end{equation}
which will be used in the analytical derivation below\cite{B} (in our
numerical work, we also use algebraic $\alpha $'s close to $1$). For an
energy $\varepsilon \gg 1$, where the relevant quantities can be described
semiclassically, the level $i$ nearest to (and below) $\varepsilon $ will be
characterized, in general, by a different pair $\left( n_{x},n_{y}\right) $
for each $\alpha $. On the average, the level number ${\cal N}$ of level $%
i=\left( n_{x},n_{y}\right) $ is 
\begin{equation}
\left\langle {\cal N}\right\rangle =\varepsilon \gg 1  \label{N_calligr}
\end{equation}
In what follows, the variations in ${\cal N}$ with $\alpha $ are not
important. Consequently, we drop $\left\langle {}\right\rangle $ and
identify ${\cal N}$ with $\varepsilon $.

The matrix element of the orbital angular momentum between the levels $%
i=\left( n_{x},n_{y}\right) $ and $f=\left( k_{x},k_{y}\right) $ ($%
\varepsilon _{f}>\varepsilon _{i}$ for definiteness) is given by 
\begin{equation}
\left| \widehat{L}_{if}\right| ^{2}=\frac{16}{\pi ^{4}}\left[ 
\begin{array}{c}
\frac{n_{x}^{2}}{\alpha }\left( \frac{1}{k_{x}+n_{x}}+\frac{1}{k_{x}-n_{x}}
\right) ^{2}\left( \frac{1}{\left( k_{y}+n_{y}\right) ^{2}}-\frac{1}{\left(
k_{y}-n_{y}\right) ^{2}}\right) ^{2} \\ 
+\alpha n_{y}^{2}\left( \frac{1}{k_{y}+n_{y}}+\frac{1}{k_{y}-n_{y}}\right)
^{2}\left( \frac{1}{\left( k_{x}+n_{x}\right) ^{2}}-\frac{1}{\left(
k_{x}-n_{x}\right) ^{2}}\right) ^{2} \\ 
-2n_{x}n_{y}\left( \frac{1}{k_{x}+n_{x}}+\frac{1}{k_{x}-n_{x}}\right) \left( 
\frac{1}{k_{y}+n_{y}}+\frac{1}{k_{y}-n_{y}}\right) \\ 
\times \left( \frac{1}{\left( k_{x}+n_{x}\right) ^{2}}-\frac{1}{\left(
k_{x}-n_{x}\right) ^{2}}\right) \left( \frac{1}{\left( k_{y}+n_{y}\right)
^{2}}-\frac{1}{\left( k_{y}-n_{y}\right) ^{2}}\right)
\end{array}
\right]  \label{Lif}
\end{equation}
where $\left( k_{x}-n_{x}\right) $ and $\left( k_{y}-n_{y}\right) $ are odd.
Retaining only the terms that contain $\left( k_{x}-n_{x}\right) $ and $%
\left( k_{y}-n_{y}\right) $, we find a simplified form 
\begin{equation}
\left| \widehat{L}_{if}\right| ^{2}\approx \frac{16}{\pi ^{4}}\left[ \frac{1%
}{\left( k_{x}-n_{x}\right) \left( k_{y}-n_{y}\right) }\left( \frac{\alpha
^{-1/2}n_{x}}{k_{y}-n_{y}}-\frac{\alpha ^{1/2}n_{y}}{k_{x}-n_{x}}\right)
\right] ^{2}  \label{Lif_simpl}
\end{equation}
For levels $i$ and $f$ that are nearest in energy 
\begin{equation}
k_{x}^{2}+k_{y}^{2}\approx n_{x}^{2}+n_{y}^{2}  \label{epsi_epsf}
\end{equation}
and for small $\left( k_{x}-n_{x}\right) $ and $\left( k_{y}-n_{y}\right) $,
we find 
\begin{equation}
n_{x}\left( k_{x}-n_{x}\right) \approx -n_{y}\left( k_{y}-n_{y}\right)
\label{epsi_epsf-2}
\end{equation}
Consequently, for $\alpha \sim 1$, 
\begin{equation}
\left| \widehat{L}_{if}\right| ^{2}\approx \frac{64}{\pi ^{4}}\frac{n_{x}^{2}%
}{\left( k_{x}-n_{x}\right) ^{2}\left( k_{y}-n_{y}\right) ^{4}}
\label{Lif_simpl-2}
\end{equation}

Clearly, the magnitude of $\left| \widehat{L}_{if}\right| ^{2}$ is
determined by the hierarchy of values $\left| k_{x}-n_{x}\right| $, $\left|
k_{y}-n_{y}\right| $ such that $%
%TCIMACRO{\limfunc{sgn}}
%BeginExpansion
\mathop{\rm sgn}%
%EndExpansion
\left( k_{x}-n_{x}\right) =-%
%TCIMACRO{\limfunc{sgn}}
%BeginExpansion
\mathop{\rm sgn}%
%EndExpansion
\left( k_{y}-n_{y}\right) $. Fig. 1 shows $\left| \widehat{L}_{if}\right|
^{2}$ as a functions of ${\cal N}$ for a single aspect ratio $\alpha
=8/31^{0.6}\approx 1.01923$. The top straight line corresponds to $%
k_{x}-n_{x}=\pm 1$, $k_{y}-n_{y}=\mp 1$ The other two lines correspond,
respectively, to $k_{x}-n_{x}=\pm 1$, $k_{y}-n_{y}=\mp 3$ and $%
k_{x}-n_{x}=\pm 1$, $k_{y}-n_{y}=\mp 5$ and $x\leftrightarrow y$
permutations. The slopes of other lines are too small for them to be visible
in this plot. Using eqs. (\ref{enn_reduced-1}) and (\ref{N_calligr}), we
have $n_{x}^{2}+n_{y}^{2}\approx 4{\cal N}/\pi .$ Combining this with (\ref
{epsi_epsf-2}) and substituting into eq. (\ref{Lif_simpl-2}), we find 
\begin{equation}
\left| \widehat{L}_{if}\right| ^{2}\approx \frac{128{\cal N}}{\pi ^{5}}
\times \left( 
\begin{array}{c}
1 \\ 
\frac{1}{45} \\ 
\frac{1}{325}
\end{array}
\right) \left( 
\begin{array}{c}
\text{for }\pm 1,\mp 1 \\ 
\text{for }\pm 1,\mp 3\text{ and }\mp 1,\pm 3 \\ 
\text{for }\pm 1,\mp 5\text{ and }\mp 1,\pm 5
\end{array}
\right)  \label{Lif_straight}
\end{equation}
for the three lines shown in Fig. 1. As seen from the figure, these straight
lines are in excellent agreement with the numerical results.

We emphasize that Fig. 1 should be understood as follows: as $i$ moves up,
from one level to the next, $\left| \widehat{L}_{if}\right| ^{2}$ ''jumps''
- up or down - between the points on straight lines (whose total number is
of order ${\cal N}$, with only three shown here), indicating orders of
magnitude fluctuations as a function of the position in the spectrum.

\subsection{Spectral average of $\left| \widehat{L}_{if}\right| ^{2}$}

It follows from (\ref{enn_reduced}) that the greater the range of $\alpha $,
the greater is the spectral $\left( x,y\right) $ asymmetry and the higher in
spectrum it is necessary to move to eliminate it\footnote{%
In other words, changes of $n_{x}$ vs. changes of $n_{y}$ lead to disparate
changes of energy for rectangles whose sides are substantially different.
However, this anisotropy is eliminated for an algebraic aspect ratio once
above a sufficiently high energy in the spectrum.}. On the other hand,
keeping $\alpha $'s close to a fixed value introduces a problem of sampling,
which, again, requires higher energies to be considered numerically.%
\footnote{%
That is, at low energies the spectra for different $\alpha $'s are very
close to each other.} Thus, there exists an inherent technical difficulty
with ensemble averaging in a box. As a practical matter, we take up to $400$
values of $\alpha \in \left[ 1,2\right] $ for the energy range $\varepsilon
\sim 10^{4}-10^{7}$. This choice of parameters proved suitable for ensemble
averaging in a numerical part of the study of statistical properties of the
energy spectrum itself.\cite{WGS} However, ensemble averaging with these
parameters does not lead to the numerical convergence of $\left\langle
\left| \widehat{L}_{if}\right| ^{2}\right\rangle $.

A portent of this can be already glimpsed from the orders of magnitude
fluctuations in $\left| \widehat{L}_{if}\right| ^{2}$ discussed in the
preceding subsection. Moreover, if ${\cal N}$ is fixed instead but different
aspect ratios are considered, the fluctuations of $\left| \widehat{L}
_{if}\right| ^{2}$ are as large as a function of $\alpha $ as they are as a
function of ${\cal N}$. Consequently, together with $\alpha $-averaging, an
additional spectral averaging over the energy interval $\left[ \varepsilon
-E/2,\varepsilon +E/2\right] $ , $E\ll \varepsilon $, is performed to
achieve numerical convergence. The interval width is taken to be $\gtrsim 
\sqrt{\varepsilon }$, which, for the above parameters, increases the
sampling range by several orders of magnitude. Interestingly, it also turns
out that spectral averaging is amenable to an analytical derivation, which
we proceed to outline here.

The key idea in this derivation is that the average of $\left| \widehat{L}
_{if}\right| ^{2}$ is dominated by the contribution from the top line in
Fig. 1. Numerically, it is clearly seen from Fig.2 where the higher set of
points corresponds to averaging over all pairs $\left( i,f\right) $ (all
lines in Fig. 1), while the lower set corresponds to averaging only over the
top line in Fig. 1($k_{x}-n_{x}=\pm 1$, $k_{y}-n_{y}=\mp 1$). The fitting
lines for the two sets are given, respectively, by 
\begin{equation}
\left\langle \left| \widehat{L}_{if}\right| ^{2}\right\rangle _{all}=0.177%
\sqrt{{\cal N}}  \label{Lif_h}
\end{equation}
and 
\begin{equation}
\left\langle \left| \widehat{L}_{if}\right| ^{2}\right\rangle _{top}=0.172%
\sqrt{{\cal N}}  \label{Lif_l}
\end{equation}
which are remarkably close. Note that we used a combined average over $400$
values of $\alpha \in \left[ 1,2\right] $ and over energy window $E=4\times
10^{4}$. Great sensitivity to the spectral averaging is obvious from Fig. 3
where the interval was reduced to $E=4\times 10^{2}$.

Physical interpretation of these results is as follows. As has been
mentioned in the preceding subsection, at a spectral point ${\cal N}$ there
are $\sim {\cal N}$ lines whose slopes can be determined (for small $\left|
k_{x,y}-n_{x,y}\right| $) via a procedure that resulted in eq. (\ref
{Lif_straight}). For a given $\alpha $, as $i$ moves upward from level to
level (that is, as ${\cal N}$ is increased), $\left| \widehat{L}_{if}\right|
^{2}$ ''jumps'' between these straight lines. In principle, a ''jump'' can
be from any one of these lines to any other. However, in determining the
average, one must remember that not only the slope decreases rapidly from
the top line down, but the probability of being on a line also decreases
rapidly from $\sim {\cal N}^{-1/2}$ on the top line to $\sim {\cal N}^{-1}$
over the course of $\sqrt{{\cal N}}$ lines$.$ Given this, and in view of the
numerical evidence above, we approximate $\left\langle \left| \widehat{L}%
_{if}\right| ^{2}\right\rangle $ by the contribution from the top line in
Fig. 1 alone. From (\ref{Lif_straight}), we find 
\begin{eqnarray}
\left\langle \left| \widehat{L}_{if}\right| ^{2}\right\rangle &=&\frac{128%
{\cal N}}{\pi ^{5}}P\left( k_{x}-n_{x}=\pm 1,k_{y}-n_{y}=\mp 1\right)
\label{Lif_av_top} \\
&=&\frac{128{\cal N}}{\pi ^{5}}P\left( k_{x}-n_{x}=\pm 1\right) P\left(
k_{y}-n_{y}=\mp 1\mid k_{x}-n_{x}=\pm 1\right)  \label{Lif_av_top-2}
\end{eqnarray}
Here $P\left( k_{x}-n_{x}=\pm 1,k_{y}-n_{y}=\mp 1\right) $ and $P\left(
k_{y}-n_{y}=\mp 1\mid k_{x}-n_{x}=\pm 1\right) $ are, respectively, the
probability that the nearest energy level pair satisfies the condition $%
\left( i,f\right) =\left( k_{x}-n_{x}=\pm 1,k_{y}-n_{y}=\mp 1\right) \ $and
a conditional probability that $k_{y}-n_{y}=\mp 1$, given that $%
k_{x}-n_{x}=\pm 1$. In what follows, these are evaluated in a series of
consecutive steps.

First, we derive the probability distribution function $p\left( n_{x}\right) 
$. Consider the energy interval defined, in accordance with (\ref
{enn_reduced-1}), by the numbers $N_{2}>N_{1}\gg 1$. A narrow interval is
defined as such that 
\begin{equation}
\delta \equiv \frac{\triangle N}{N}=\frac{\triangle {\cal N}}{{\cal N}}\ll 1
\label{delta}
\end{equation}
where $N$ is the center of the interval and $\triangle N$ is its width: 
\begin{equation}
N=\frac{N_{2}+N_{1}}{2}=\frac{4}{\pi }\varepsilon =\frac{4}{\pi }{\cal N}%
\text{, }\triangle N=N_{2}-N_{1}=\frac{4}{\pi }E=\frac{4}{\pi }\triangle 
{\cal N}  \label{DelataN}
\end{equation}
All pair points $\left( n_{x},n_{y}\right) $ in this interval are located
between the two quarter-circles shown in Fig. 4, 
\begin{equation}
N_{1}\leq n_{x}^{2}+n_{y}^{2}\leq N_{2}\text{, }n_{x,y}>0  \label{area}
\end{equation}
We propose a simple ansatz whereby $p\left( n_{x}\right) dn_{x}$ is just the
number of states in the interval $dn_{x}$, as illustrated by the shaded
areas, which gives 
\begin{equation}
p\left( n_{x}\right) =\frac{4}{\pi \triangle N}\times \left( 
\begin{array}{c}
\sqrt{N_{2}-n_{x}^{2}}-\sqrt{N_{1}-n_{x}^{2}} \\ 
\sqrt{N_{2}-n_{x}^{2}}
\end{array}
\right) \left( 
\begin{array}{c}
0\leq n_{x}\leq \sqrt{N_{1}} \\ 
\sqrt{N_{1}}\leq n_{x}\leq \sqrt{N_{2}}
\end{array}
\right)  \label{p_nx}
\end{equation}
Fig. 5 shows this formula vis-a-vis the numerical evaluation for the same $%
\alpha $ as in Fig. 1 and for $0.9\times 10^{5}\leq \varepsilon \leq
1.1\times 10^{5}$. Clearly, the two are in excellent agreement.

Next, we evaluate the distribution function $p\left( k_{x}-n_{x}\right) $ by
convoluting $p\left( n_{x}\right) $ and $p\left( k_{x}\right) $ under the
assumption of no correlations between $n_{x}$ and $k_{x}$: 
\begin{equation}
p\left( k_{x}-n_{x}\right) =\int_{0}^{\sqrt{N_{2}}}p\left(
k_{x}-n_{x}+t\right) p\left( t\right) dt  \label{p_kxnx}
\end{equation}
The resulting formula is very complicated and in Fig. 6 we will only show
its plot vis-a-vis numerical evaluation for $0.99\times 10^{6}\leq
\varepsilon \leq 1.01\times 10^{6}$ for a single $\alpha $ and for an
ensemble average over $100$ algebraic $\alpha \in \left[ 1,1.25\right] $.
Clearly the agreement between the two is very good. Notice that the dip at
zero in the numerical distribution function is due to the fact that the
probability of $k_{x,y}-n_{x,y}=0$ is suppressed for the nearest levels
(see, for instance eq. (\ref{epsi_epsf-2}) showing that $\left(
k_{x}-n_{x}\right) $ and $\left( k_{y}-n_{y}\right) $ should be finite and
of opposite signs). Notice also that we treat $\left( k_{x}-n_{x}\right) $
as a continuous variable and so a deviation from the numerical results
should be anticipated for small discrete values of this variable.

As is clear from (\ref{Lif_av_top-2}), we only need the probability $P\left(
k_{x}=n_{x}\pm 1\right) $ for our purposes. However, for the reasons just
mentioned, we cannot precisely determine it from the distribution obtained
via (\ref{p_kxnx}) and shown in Fig. 6 by solid line. Consequently, we use
an ansatz where this probability is approximated by the area of width$\ \sim
1$ near the maximum of the distribution, that is 
\begin{equation}
P\left( k_{x}-n_{x}=\pm 1\right) \approx p\left( 0\right)  \label{Pp}
\end{equation}
where 
\begin{equation}
p\left( 0\right) =\int_{0}^{\sqrt{N_{2}}}p^{2}\left( t\right) dt=\frac{32}{%
3\pi ^{2}\sqrt{N_{2}}}\frac{1+\left( 1-\delta \right) ^{3/2}+\delta K\left(
1-\delta \right) -\left( 2-\delta \right) E\left( 1-\delta \right) }{\delta
^{2}}  \label{p_0}
\end{equation}
and $K$ and $E$ are the elliptic functions. Expanding this expression for
small $\delta $, while simultaneously replacing $N_{2}$ by $N$, we find 
\begin{equation}
p\left( 0\right) \approx \frac{3+8\ln 2-2\ln \delta }{\pi ^{2}\sqrt{N}}
\label{p_0-expand}
\end{equation}
Fig. 7a shows numerical results for $P\left( k_{x}-n_{x}=\pm 1\right) $
vis-a-vis eqs. (\ref{p_0}) and (\ref{p_0-expand}). Fig. 7b also shows $%
P\left( k_{x}-n_{x}=\pm 1\right) \sqrt{{\cal N}}$ vis-a-vis $p\left(
0\right) \sqrt{{\cal N}}$. The latter eliminates the main $\sqrt{{\cal N}}$
dependence and shows that the remainder is a slow-growing function of ${\cal %
N}$. Unfortunately, due to indeterminacy inherently present in our ansatz,
it is incapable of exactly describing this function.

To evaluate the conditional probability $P\left( k_{y}-n_{y}=\mp 1\mid
k_{x}-n_{x}=\pm 1\right) $, we notice that by substituting $k_{x}-n_{x}=-1$
into (\ref{epsi_epsf}) and using (\ref{enn_reduced-1}) one finds\footnote{%
Evaluating specifically $P\left( k_{y}-n_{y}=1\mid k_{x}-n_{x}=-1\right) $.} 
\begin{equation}
k_{y}-n_{y}=\sqrt{N-\left( n_{x}-1\right) ^{2}}-\sqrt{N-n_{x}^{2}}
\label{kyny_nx}
\end{equation}
whereof 
\begin{equation}
n_{x}=\frac{1}{2}\left[ 1+\left( k_{y}-n_{y}\right) \sqrt{\frac{4N-\left(
k_{y}-n_{y}\right) ^{2}}{1+\left( k_{y}-n_{y}\right) ^{2}}}\right]
\label{nx_kyny}
\end{equation}
in the limit $N\gg 1$. Consequently, using the distribution function (\ref
{p_nx}), we find the distribution function for $p\left( k_{y}-n_{y}\mid
k_{x}-n_{x}=-1\right) $ as 
\begin{equation}
p\left( k_{y}-n_{y}\mid k_{x}-n_{x}=-1\right) =p\left( n_{x}\left(
k_{y}-n_{y}\right) \right) \frac{dn_{x}}{d\left( k_{y}-n_{y}\right) }
\label{p_kyny}
\end{equation}
where is $n_{x}\left( k_{y}-n_{y}\right) $ is a function given by eq. (\ref
{nx_kyny}). Combining now eqs. (\ref{kyny_nx})-(\ref{p_kyny}) with eq. ((\ref
{p_nx}) we find the analytical form (not shown here due to its complexity)
of the conditional probability distribution function and plot it in Fig. 8
vis-a-vis the numerical results.

Before discussing analytical vs. numerical results, we wish to pause briefly
on the slight difference between $p\left( k_{y}-n_{y}\mid
k_{x}-n_{x}=-1\right) $ and $p\left( \left| k_{y}-n_{y}\right| \mid
k_{x}-n_{x}=1\right) $, as seen in Fig. 8. It follows from a more careful
examination of eq. (\ref{epsi_epsf}). Indeed, it should be written as 
\begin{equation}
k_{x}^{2}+k_{y}^{2}\approx n_{x}^{2}+n_{y}^{2}+\widetilde{\Delta }
\label{epsi_epsf_mod}
\end{equation}
where $\widetilde{\Delta }>0$ is the separation between the levels.
Consequently, (\ref{epsi_epsf-2}) should be replaced by 
\begin{equation}
n_{x}\left( k_{x}-n_{x}\right) \approx -n_{y}\left( k_{y}-n_{y}\right) +%
\frac{\widetilde{\Delta }}{2}  \label{epsi_epsf_mod-2}
\end{equation}
Obviously then, it is possible to have $k_{y}-n_{y}=0$ when $k_{x}-n_{x}=1$
but not when $k_{x}-n_{x}=-1$. This is reflected in numerical curves in Fig.
8 where the conditional probability $p\left( k_{y}-n_{y}\mid
k_{x}-n_{x}=-1\right) $ does not have a value at $k_{y}-n_{y}=0$ while $%
p\left( \left| k_{y}-n_{y}\right| \mid k_{x}-n_{x}=1\right) $ does.

We now turn to a closer analysis of Fig. 8. Obviously, the numerical
distribution is much broader than the analytical distribution. The reason
for that is that the former was obtained using a large number of $\alpha $
-values whereas the latter was obtained in the approximation where $\alpha $
was set to $\sim 1$ and hence (\ref{enn_reduced-1}). Clearly, even if the $%
\alpha $-values are sufficiently close to $1$, as in our simulations,
different ensembles are nonetheless sampled for different $\alpha $'s (for
sufficiently high energies, as has been previously mentioned) using the
exact formula (\ref{enn_reduced}). As a result, the likelihood of larger $%
\left| k_{y}-n_{y}\right| $, given that $\left| k_{x}-n_{x}\right| =1$, is
higher in this case than in our analytical derivation based on (\ref
{enn_reduced-1}).

Of note, however, is that for small $\left| k_{y}-n_{y}\right| $ the
analytical curve is well approximated by 
\begin{equation}
p\left( \left| k_{y}-n_{y}\right| \mid \left| k_{x}-n_{x}\right| =1\right)
\approx \frac{2}{\pi \left( 1+\left| k_{y}-n_{y}\right| ^{2}\right) }+\frac{%
\Delta N}{2\pi N}\stackrel{\left| k_{y}-n_{y}\right| =0}{\rightarrow }\frac{2%
}{\pi }+\frac{\Delta N}{2\pi N}  \label{p_kyny-expand}
\end{equation}
We assume that, up to a constant, this gives a good approximation to the
dependence on $\Delta N/N=\Delta {\cal N}/{\cal N}.$ Consequently, we
propose, similarly to $p\left( k_{x}-n_{x}\right) $ before, that $P\left(
k_{y}-n_{y}=\mp 1\mid k_{x}-n_{x}=\pm 1\right) $ can be approximated by 
\begin{equation}
P\left( k_{y}-n_{y}=\mp 1\mid k_{x}-n_{x}=\pm 1\right) \approx const\times
p\left( 0\mid \left| k_{x}-n_{x}\right| =1\right)  \label{P1p}
\end{equation}
Fig. 9 shows a plot with an empirical $const=.225$ used both in ''exact'' $%
p\left( 0\mid \left| k_{x}-n_{x}\right| =1\right) $, obtained from (\ref
{p_kyny}), and approximation (\ref{p_kyny-expand}) vis-a-vis the numerical
result. Clearly, it provides credence to our ansatz.

\subsection{Summary}

The key results of this section are summarized in Figs. 1 and 2 and eqs. (%
\ref{Lif_h}) and (\ref{Lif_l}):

\begin{itemize}
\item  Given the energy (the level number) $\simeq {\cal N}$, the magnitude
of $\left| \widehat{L}_{if}\right| ^{2}$ will fall on one of $\sim {\cal N}$
lines, such as the three straight line shown in Fig. 1 and given by eqs. (%
\ref{Lif_straight}). Which line specifically it will be on depends on a
particular aspect ratio of the rectangle $\alpha $. Conversely, for a given $%
\alpha $, the line it will be on depends on a specific nearest levels pair $
\left( i,f\right) $ in the vicinity of the energy considered. In other
words, $\left| \widehat{L}_{if}\right| ^{2}$ experiences orders of magnitude
fluctuation both as a function of $\alpha $ and as a function of energy.

\item  For numerical convergence it was necessary to perform a combined
ensemble averaging (over $\alpha $) and energy averaging (over energy
interval $\triangle {\cal N}\ll {\cal N}$). The result is given by eq. (\ref
{Lif_h}) and its magnitude almost entirely derives from the contribution of
the top straight line in Fig. 1, $\left| \widehat{L}_{if}\right| ^{2}=128%
{\cal N}/\pi ^{5}$, as seen by comparison with (\ref{Lif_l}). The
probability to find $\left| \widehat{L}_{if}\right| ^{2}$ on the top line is 
$\propto {\cal N}^{-1/2}$, which explains why $\left\langle \left| \widehat{L%
}_{if}\right| ^{2}\right\rangle \propto \sqrt{{\cal N}}$. It is central to
our derivation that the conditional probability $P\left( k_{y}-n_{y}=\mp
1\mid k_{x}-n_{x}=\pm 1\right) $ only weakly depends on ${\cal N}$, $\sim
0.14\left( 1+\Delta {\cal N}/4{\cal N}\right) $.\newpage
\end{itemize}

\section{Orbital susceptibility of a free-electron gas}

\subsection{Introduction}

It was recently proposed that, at zero temperature the orbital magnetic
susceptibility, of free electrons in disordered systems can be explained by
the two level van Vleck response that involves the last occupied (Fermi)
level and the first unoccupied level\cite{S1},\cite{S2}. Whereas for an
occasional Fermi level in a given realization of disorder, or for an
occasional realization of disorder given a Fermi level close to a fixed
energy value, the response can be diamagnetic, in the vast majority of cases
it is paramagnetic. This prediction was recently verified numerically in
Ref. \cite{GB}, which also confirmed that both the mean susceptibility and
the susceptibility distribution function (mesoscopic fluctuations) can be
quite accurately described by the two level model\cite{S1},\cite{S2}.

Orbital susceptibility of rectangles has been previously studied both
analytically and numerically \cite{vRvL}-\cite{RUJ}. However, analytically
the approach that generally works only at sufficiently high temperature was
used and numerically only the rational aspect ratios of rectangles were
examined (e.g. squares). Conversely, here we are concerned with a strictly $%
T=0$ response for the irrational (algebraic in this numerical evaluation)
aspect ratios. Our main conclusion is that, in complete analogy with the
disordered systems, both the average and the fluctuations are successfully
described by the two level van Vleck response that involves the Fermi level
and the first unoccupied level. Furthermore, the largest contribution to the
response comes from the values $\left| \widehat{L}_{if}\right| ^{2}$ from
the top straight line in Fig. 1, as explained in the preceding Section,
while the largest fluctuation occur when, for the points on that line, the
energy difference $\varepsilon _{f}-\varepsilon _{i}$ between the nearest
levels is particularly small.

It will be observed that the orbital susceptibility exhibits a striking
absence of self-averaging. While in part it is due to the fact that we
evaluate the zero-field response at zero temperature, the underlying physics
underscores, nonetheless, that mesoscopic effects are much more pronounced
in classically integrable than in classically chaotic systems.

\subsection{Orbital susceptibility of a rectangular box}

Using the dimensionless notations, where the energy is measured in units of $%
\Delta $ and the susceptibility in units of $\mu _{B}^{2}/\Delta $, we find
the total orbital susceptibility as follows: 
\begin{equation}
\chi _{tot}=-\sum_{i=1}^{{\cal N}}\frac{2\pi \left\langle i\right|
x^{2}+y^{2}\left| i\right\rangle }{A}+\sum_{i=1}^{{\cal N}}\sum_{f={\cal N}%
+1}^{{\cal \infty }}\frac{2\left| \widehat{L}_{if}\right| ^{2}}{\varepsilon
_{f}-\varepsilon _{i}}  \label{khi_tot}
\end{equation}
where $\varepsilon _{i}$, $i=\left( n_{x},n_{y}\right) $, is the unperturbed
(zero-field) spectrum (\ref{enn_reduced}). While the Landau gauge was used
in this expression, the final result is gauge-independent (for a discussion,
see Ref.\cite{GB}). The diamagnetic matrix elements are easily calculated
and are given by 
\begin{equation}
\left\langle i\right| x^{2}+y^{2}\left| i\right\rangle =\frac{A}{12}\left[
\alpha ^{1/2}\left( 1-\frac{6}{\pi n_{x}^{2}}\right) +\alpha ^{-1/2}\left( 1-%
\frac{6}{\pi n_{y}^{2}}\right) \right]  \label{diam}
\end{equation}

Fig. 10 shows the result of numerical evaluation of $\left\langle \chi
_{tot}\right\rangle $ plotted as a function of ${\cal N}$, vis-a-vis $%
\left\langle 2\left| \widehat{L}_{if}\right| ^{2}\left( \varepsilon
_{f}-\varepsilon _{i}\right) ^{-1}\right\rangle $, where $\varepsilon _{i}$
and $\varepsilon _{f}$ are now limited to the Fermi level and the first
unoccupied level, $i={\cal N},$ $f={\cal N}+1$. This is motivated by the
surmise that the contributions of the two sums in eq. (\ref{khi_tot}) -
diamagnetic and paramagnetic - largely cancel each other over the Fermi sea
and the total susceptibility, on average, can be explained by a single term
in the van Vleck sum, namely, the one between the last occupied and the
first unoccupied levels. (The analogous surmise in disordered systems\cite
{S1},\cite{S2} had been already verified numerically\cite{GB}.) The subset
of the latter, $\left\langle 2\left| \widehat{L}_{if}\right| ^{2}\left(
\varepsilon _{f}-\varepsilon _{i}\right) ^{-1}\right\rangle _{top}$, with
the contributions only from the top line in Fig. 1 is also shown. While the
difference in the distribution function of $\left( \varepsilon
_{f}-\varepsilon _{i}\right) $ on the top line relative to the Poissonian
should be noted (and will be discussed in a separate publication), the
dominance of the $k_{x}-n_{x}=\pm 1,k_{y}-n_{y}=\mp 1$ contribution to $%
\left| \widehat{L}_{if}\right| $ suggests that the top line should dominate
the contribution from the $i={\cal N},$ $f={\cal N}+1$ term also. Finally, $%
2\left\langle \left| \widehat{L}_{if}\right| ^{2}\right\rangle \left\langle
\left( \varepsilon _{f}-\varepsilon _{i}\right) ^{-1}\right\rangle $, is
also shown in Fig. 10, where $\left\langle \left| \widehat{L}_{if}\right|
^{2}\right\rangle $ is shown in Fig. 2 and approximated by eq. (\ref{Lif_h})
and $\left\langle \left( \varepsilon _{f}-\varepsilon _{i}\right)
^{-1}\right\rangle $ is evaluated numerically and found to be (in units of $%
\Delta ^{-1}$) 
\begin{equation}
\left\langle \frac{1}{\varepsilon _{f}-\varepsilon _{i}}\right\rangle
\approx 15.5  \label{invspacing_ave}
\end{equation}

Despite a combined averaging over $400$ $\alpha $'s and energy interval $%
4000 $-wide, the absence of self averaging is still evident in this plot.
(To further emphasize the predominance of large fluctuations, in Fig. 11 we
show the same $\left\langle \chi _{tot}\right\rangle $ as in Fig. 10
vis-a-vis $\left\langle \chi _{tot}\right\rangle $ that was obtained for the
same $\alpha $-ensemble but whose energy averaging was performed over
intervals $10 $ times narrower.) On the other hand, our surmise that the two
level van Vleck paramagnetism accurately describes the average orbital
response is evident from Fig. 10 as the structure of $\left\langle \chi
_{tot}\right\rangle $ vs. ${\cal N}$ is well reproduced by the nearest level
contributions alone; the difference between the two are the contributions
from the terms in the double-sum of (\ref{khi_tot}) that are due to the
levels further below and above the Fermi level.

The large value of $\left\langle \left( \varepsilon _{f}-\varepsilon
_{i}\right) ^{-1}\right\rangle $ in (\ref{invspacing_ave}) is readily
understood from the exponential (Poissonian) distribution function of the
level spacings\cite{G},\cite{BFFMPW}, $p\left( \varepsilon _{f}-\varepsilon
_{i}\right) =\exp \left[ -\left( \varepsilon _{f}-\varepsilon _{i}\right)
\right] $, whereof 
\begin{equation}
\left\langle \frac{1}{\varepsilon _{f}-\varepsilon _{i}}\right\rangle
=\int_{\epsilon }^{\infty }\frac{\exp \left( -x\right) }{x}dx=\ln \frac{1}{%
\epsilon }  \label{diverg}
\end{equation}
where $\epsilon $ is a cut-off. Since we evaluate the zero field response at
zero temperature, the cut-off is the smallest spacing observed for the
values of $\alpha $ and energies considered here, which happens to be $\sim
10^{-7}$ (the mean level spacing being $1$). As already mentioned, the
distribution function of the level spacings on the top line of Fig. 1 will
be discussed elsewhere, however, it turns out that $\left\langle \left(
\varepsilon _{f}-\varepsilon _{i}\right) ^{-1}\right\rangle $ is also
formally divergent and, for the parameters used here, numerical evaluation
gives $\left\langle \left( \varepsilon _{f}-\varepsilon _{i}\right)
^{-1}\right\rangle \approx 12.5$. It should be pointed out however that in
reality, even at zero temperature, the magnetic field itself introduces a
natural cut-off; for disordered systems this was discussed in Ref.\cite{S2}.

We now turn to the study of fluctuations. To understand the nature of
mesoscopic fluctuations, we investigate the energy differences between the
nearest levels. We limit our consideration to the top line in Fig. 1 as it
dominates the response, both in terms of the mean and the fluctuations.
Figs. 12-14 show plots of $\left( \varepsilon _{f}-\varepsilon _{i}\right)
^{-1}$ and $\varepsilon _{f}-\varepsilon _{i}$ as a function of ${\cal N}$
for three particular values of $\alpha $, respectively $\alpha
_{1}=457^{0.855}/185\approx 1.01638$, $\alpha _{2}=911^{0.755}/150\approx
1.14379$, and $\alpha _{3}=643^{0.655}/63\approx 1.09657$. First, we notice
the existence of very large values of $\left( \varepsilon _{f}-\varepsilon
_{i}\right) ^{-1}$, which are the source of very large mesoscopic
fluctuations. Second, we notice patterns in the structure of $\left(
\varepsilon _{f}-\varepsilon _{i}\right) ^{-1}$ and $\varepsilon
_{f}-\varepsilon _{i}$ as a function of ${\cal N}$. Lastly, we notice that
the size of the peaks and the pattern structure is very strongly $\alpha $%
-dependent.

It turns out that the patterns in Figs. 12-14,as well as the size and the
location of the peaks, can be determined analytically. Considering for
simplicity the case of $\alpha =1+\beta $, $0<\beta <1/2$, we find 
\begin{equation}
\frac{4\alpha ^{1/2}}{\pi }\left( \varepsilon _{f}-\varepsilon _{i}\right) = 
%TCIMACRO{
%\binom{2\left( n_{x}-n_{y}+1\right) -\beta \left( 2n_{y}-1\right) }{2\left( n_{y}-n_{x}+1\right) +\beta \left( 2n_{y}+1\right) } }
%BeginExpansion
{2\left( n_{x}-n_{y}+1\right) -\beta \left( 2n_{y}-1\right)  \choose 2\left( n_{y}-n_{x}+1\right) +\beta \left( 2n_{y}+1\right) }%
%EndExpansion
%TCIMACRO{
%\binom{k_{x}-n_{x}=1,k_{y}-n_{y}=-1}{k_{x}-n_{x}=-1,k_{y}-n_{y}=1} }
%BeginExpansion
{k_{x}-n_{x}=1,k_{y}-n_{y}=-1 \choose k_{x}-n_{x}=-1,k_{y}-n_{y}=1}%
%EndExpansion
\label{epsi_epsf_top}
\end{equation}
The algebraic $\beta $ can be represented as a series 
\begin{equation}
\beta =\frac{1}{p}+\frac{1}{q}+\frac{1}{r}+\ldots  \label{beta}
\end{equation}
where $p>2$, $\left| q\right| >p^{2}$, $\left| r\right| >q^{2}$ ,$\ldots $
are integers. The series can be truncated 
\begin{eqnarray}
\beta &=&\frac{1}{p}+\eta  \nonumber \\
\beta &=&\frac{1}{p}+\frac{1}{q}+\eta  \label{beta_truncated} \\
\beta &=&\frac{1}{p}+\frac{1}{q}+\frac{1}{r}+\eta  \nonumber \\
&&\ldots  \nonumber
\end{eqnarray}
where $\eta $ is the residual algebraic number. For $\beta $'s corresponding
to Figs. 12-14, the truncated series are as follows: 
\begin{eqnarray}
\beta _{1} &=&\frac{1}{61}-\frac{1}{81672}-5.1\times 10^{-11}  \nonumber \\
\beta _{2} &=&\frac{1}{7}+\frac{1}{1076}+\frac{1}{2616460}+5.6\times 10^{-13}
\label{beta_examples} \\
\beta _{3} &=&\frac{1}{10}-\frac{1}{292}-\frac{1}{472427}\times 10^{-13} 
\nonumber
\end{eqnarray}
Specifics of truncation depend on the position in the energy spectrum, to
which numerical analysis is extended, and on the values of integers $p$, $q$%
, $r$, $\ldots $, as explained below.

The key to understanding the spacings structure in Figs. 12-14 is that it
can be completely described by as few as the first two rational numbers in
the approximation of $\beta $ (\ref{beta_truncated}). (Only for the very
rare occurrences of degeneracy, $\varepsilon _{f}-\varepsilon _{i}=0$, does
the residual term $\eta $ needs to be considered.) Therefore, we turn to the
analysis of the interplay between the rational numbers. We begin with the
spacings structure for $\beta =p^{-1}$, shown in Figs. 15 for $p=100$.
Analytically, (\ref{epsi_epsf_top}) yields the following series of straight
lines, as a function of $n_{y}$: 
\begin{eqnarray}
\frac{4\alpha ^{1/2}}{\pi }\left( \varepsilon _{f}-\varepsilon _{i}\right)
&=&2C+2+\frac{1}{p}-\frac{2}{p}n_{y}\text{, } 
%TCIMACRO{
%\binom{pC+\frac{1}{2}\leq n_{y}\leq p\left( C+1\right) +\frac{1}{2}}{C=0,1,2,\ldots } }
%BeginExpansion
{pC+\frac{1}{2}\leq n_{y}\leq p\left( C+1\right) +\frac{1}{2} \choose C=0,1,2,\ldots }%
%EndExpansion
\text{, }k_{x}-n_{x}=1,k_{y}-n_{y}=-1  \nonumber \\
&&  \label{ei_ef_top_rat1} \\
\frac{4\alpha ^{1/2}}{\pi }\left( \varepsilon _{f}-\varepsilon _{i}\right)
&=&-2C+2+\frac{1}{p}+\frac{2}{p}n_{y\text{, }} 
%TCIMACRO{
%\binom{p\left( C-1\right) -\frac{1}{2}\leq n_{y}\leq pC-\frac{1}{2}}{C=1,2,3,\ldots } }
%BeginExpansion
{p\left( C-1\right) -\frac{1}{2}\leq n_{y}\leq pC-\frac{1}{2} \choose C=1,2,3,\ldots }%
%EndExpansion
\text{, }k_{x}-n_{x}=-1,k_{y}-n_{y}=1  \nonumber
\end{eqnarray}
Here $C=n_{x}-n_{y}$ is a non-negative integer, due to the choice of $\alpha
>1$, and the limits on $n_{y}$ are determined from the condition 
\begin{equation}
0\leq \frac{4\alpha ^{1/2}}{\pi }\left( \varepsilon _{f}-\varepsilon
_{i}\right) \leq 2  \label{epsi_epsf_top_cond}
\end{equation}
where the second inequality can be understood from the fact that changing $C$
by $1$ changes $4\alpha ^{1/2}\left( \varepsilon _{f}-\varepsilon
_{i}\right) /\pi $ by $2$, making it extremely unlikely that $\varepsilon
_{f}$ is the nearest level to $\varepsilon _{i}$ if the second constraint in
(\ref{epsi_epsf_top_cond}) is not satisfied; this is also confirmed
numerically. Furthermore, the maxima in Fig. 15a (minima in Fig. 15b) are
found as follows: 
\begin{eqnarray}
\frac{4\alpha ^{1/2}}{\pi }\left( \varepsilon _{f}-\varepsilon _{i}\right)
_{\min } &=&\frac{1}{p}\text{, when }%
%TCIMACRO{\binom{n_{y}=p\left( C+1\right) }{C=0,1,2,\ldots } }
%BeginExpansion
{n_{y}=p\left( C+1\right)  \choose C=0,1,2,\ldots }%
%EndExpansion
\text{, }k_{x}-n_{x}=1,k_{y}-n_{y}=-1  \nonumber \\
&&  \label{ei_ef_top_rat1-min} \\
\frac{4\alpha ^{1/2}}{\pi }\left( \varepsilon _{f}-\varepsilon _{i}\right)
&=&\frac{1}{p}\text{, when }%
%TCIMACRO{\binom{n_{y}=p\left( C-1\right) }{C=2,3,4,\ldots } }
%BeginExpansion
{n_{y}=p\left( C-1\right)  \choose C=2,3,4,\ldots }%
%EndExpansion
\text{, }k_{x}-n_{x}=-1,k_{y}-n_{y}=1  \nonumber
\end{eqnarray}
Combining eqs. (\ref{ei_ef_top_rat1}) and (\ref{ei_ef_top_rat1-min}) with 
\begin{equation}
{\cal N}=\frac{\pi \left( n_{x}^{2}+\alpha n_{y}^{2}\right) }{4\alpha ^{1/2}}
\label{N_calligr-2}
\end{equation}
completely describes the spacings structure as a function of ${\cal N}$, for
instance, as shown in Figs. 15.

We turn to the next approximation in (\ref{beta_truncated}), $\beta
=p^{-1}+q^{-1}.$ For simplicity, we will consider only $%
k_{x}-n_{x}=1,k_{y}-n_{y}=-1$ and $q<0$. In this case, the spacings
structure is described by the following equations: 
\begin{equation}
\frac{4\alpha ^{1/2}}{\pi }\left( \varepsilon _{f}-\varepsilon _{i}\right) =%
\frac{2t+2p+1}{p}+\frac{1-2n_{y}}{q}\text{, } 
%TCIMACRO{
%\binom{1\leq n_{y}\leq \frac{\left( 2t+1\right) q+p}{2p}\text{, }-p\leq t<0}{\frac{\left( 2t+2p+1\right) q+p}{2p}\leq n_{y}\leq \frac{\left( 2t+1\right) q+p}{2p}\text{, }t\leq -p-1} }
%BeginExpansion
{1\leq n_{y}\leq \frac{\left( 2t+1\right) q+p}{2p}\text{, }-p\leq t<0 \choose \frac{\left( 2t+2p+1\right) q+p}{2p}\leq n_{y}\leq \frac{\left( 2t+1\right) q+p}{2p}\text{, }t\leq -p-1}%
%EndExpansion
\label{ei_ef_top_rat2}
\end{equation}
Here $t$ is an integer and possible $n_{y}$'s are subject to the constraint 
\begin{equation}
t=pC-n_{y}  \label{constraint}
\end{equation}
where, as before, $C=n_{x}-n_{y}\geq 0$ is an integer. The minima of the
structure are found at 
\begin{equation}
n_{y}=p\left\lfloor \frac{t}{p}+\frac{\left( 2t+2p+1\right) q}{2p^{2}}+\frac{%
1}{2p}+1\right\rfloor -t\text{, }t\leq -p-1  \label{ny_min}
\end{equation}
where the $\left\lfloor {}\right\rfloor $ brackets denote the floor (integer
value) function. Together, eqs. (\ref{N_calligr-2})-(\ref{ny_min})
completely describe the spacings structure in Figs. 12. Furthermore,
generalization to the next iteration, $\beta =p^{-1}+q^{-1}+r^{-1}$, is
straightforward and accounts for Figs. 13 and 14.

Semiquantitatively, the peak structure of $\left( \varepsilon
_{f}-\varepsilon _{i}\right) ^{-1}$ along ${\cal N}$ axis is summarized
below. In what follows, we will discuss the spacings structure as a function
of $2n_{y}$, which is easily converted to that of ${\cal N}$, as explained
above. The upper limit of ${\cal N}$, vs. the values of $p$, $q$, $r$, $%
\ldots $ in approximation of a particular $\alpha $ ($\beta $), determines
how many terms is necessary to keep in (\ref{beta_truncated}). Namely, $%
q^{-1}$ enters in (\ref{beta_truncated}) when $2n_{y}\gtrsim q/p$ and $%
r^{-1} $ enters when $2n_{y}\gtrsim r/q$; this is illustrated by (\ref
{beta_examples}), as applied to Figs. 12-14. Noting that peaks correspond to
the end points (negative slope) and first points (positive slope) of the
straight lines given by eqs. (\ref{ei_ef_top_rat1}), (\ref{ei_ef_top_rat2}),
etc., and calling the distance between the tallest peaks ''period,'' we find
the following\footnote{%
It should be emphasized that not all peaks are observed for the ${\cal N}$
considered; in fact it is possible to not observe any peaks at all, in which
case the maxima of the structure are determined by the end points of the
interval on the ${\cal N}$ axis$.$}:

\begin{itemize}
\item  Peaks of $p^{-1}$: ''period'' $\sim p$; height of peaks is $p$.

\item  Peaks of $p^{-1}+q^{-1}$: (assuming that $pq$ and $p+q$ have no
common factor) period $\sim 2pq$, with $\sim p^{2}$ peaks per period
separated by distance $\sim 2q/p$; upper limit of peaks is $pq$ (heights of
peaks are $pq$, $pq/3$, $\ldots $, $\sim q/p$, or in reverse order), for $%
p+q $ odd, and $\infty $ ($\infty $, $pq/2$, $pq/4$, $\ldots ,$ $\sim q/p$)
for $p+q$ even ($\infty $ indicates level degeneracy, $\varepsilon
_{f}-\varepsilon _{i}=0$, so that the residual term $\eta $ must be
considered in (\ref{beta_truncated})).

\item  Peaks of $p^{-1}+q^{-1}+r^{-1}$: (assuming that $pqr$ and $pq+qr+pr$
have no common factor) period $\sim 2pqr$, with $\sim p^{2}q^{2}$ peaks per
period separated by distance $\sim 2r/pq$; upper limit of peaks is $pqr$
(heights of peaks $pqr$, $pqr/3$, $\ldots $, $\sim r/qp$, or in reverse
order), for $pq+qr+pr$ odd, and $\infty $ ($\infty $, $pqr/2$, $pqr/4$, $%
\ldots ,$ $\sim r/qp$) for $pq+qr+pr$ even.
\end{itemize}

\subsection{Summary}

The central result of this section is shown in Fig. 10. It demonstrates that
the mean zero-temperature, zero-field orbital magnetic susceptibility of a
free electron gas in a rectangular box can be explained in terms of a
two-level van Vleck response - that of the last occupied (Fermi)\ and the
first unoccupied levels. Furthermore, it is dominated by the contributions
from the top line in Fig. 1, namely $k_{x}-n_{x}=\pm 1,k_{y}-n_{y}=\mp 1$,
which is also the largest contributor to $\left\langle \left| \widehat{L}%
_{if}\right| ^{2}\right\rangle $ discussed in the preceding section. In
fact,the mean value of susceptibility is reasonably well described by $%
\left\langle \left| \widehat{L}_{if}\right| ^{2}\right\rangle \left\langle
\left( \varepsilon _{f}-\varepsilon _{i}\right) ^{-1}\right\rangle ,$ where $
\left\langle \left( \varepsilon _{f}-\varepsilon _{i}\right)
^{-1}\right\rangle $ is large due to the absence of correlations for small
level separations.

It is also evident that the orbital susceptibility is largely a non
self-averaging quantity, as seen from Figs. 10 and 11. This is due to the
existence of huge variations in inverse level spacings, which, in turn,
allow for such large contributions that may singularly outweigh the totality
of more typical contributions in the average response. Such variations were
explained in terms of a decomposition of algebraic aspect ratios into
rational numbers, whose interplay in (\ref{beta_truncated}) is crucial for
understanding the peaks of $\left( \varepsilon _{f}-\varepsilon _{i}\right)
^{-1}$. It must be borne in mind, however, that this feature of the orbital
susceptibility is very fragile with respect to perturbations and that
mesoscopic fluctuations will be suppressed at finite temperatures (or even
by finite values of the magnetic field); we intend to address this problem
elsewhere.

\section{Conclusions}

The most striking feature revealed here is the non-self-averaging property
of physical quantities in a rectangular box, which represents a class of
integrable billiard problems. We had previously discussed\cite{WGS} the
persisting long-range correlations in the semiclassical energy spectrum of
this system. These correlations are more complex than those in classically
chaotic (disordered) systems\cite{G}-\cite{E}. In particular, we discussed
the large, non-decaying oscillations of the level number variance on an
energy interval as a function of the interval width. Similarly, we find that
mesoscopic fluctuations here are much more pronounced than in classically
chaotic systems. For instance, while eqs. (\ref{Lif_dirt}) and (\ref
{Lif_rect}) point to the same order of magnitude in a ballistic disordered
system, $\ell \sim L$, and a rectangular billiard, the latter will have much
larger fluctuations (we have discussed the difficulties with averaging in
text).

The one similarity that stands out for both integrable and chaotic systems
is that both the average orbital susceptibility of the free electron gas and
its fluctuations can be well described by a two level van Vleck response
that couples the last occupied (Fermi) and the first unoccupied levels. For
disordered systems, this has been demonstrated previously in Refs.\cite{S1},%
\cite{S2}, and\cite{GB} and for an integrable case in this work. The
difference, however, is that in disordered systems the non-self-averaging
effects are less pronounced: in the absence of cut-offs (temperature, finite
magnetic field, etc.), the average is well defined and only the higher
cumulants are divergent. In a rectangular box, even the average is already
ill-defined, as pointed out in discussion of eq. (\ref{diverg}).

Our next step will be to investigate the effect of temperature on the
orbital magnetism of integrable systems. Towards this end, we will apply
Imry's formalism, which allows to express the average response in terms of
the level correlation function.\cite{SS} For rectangles, the latter is now
well understood, including as function of magnetic field.\cite{WGS} This
formalism works well at sufficiently high temperature and should provide an
insight into the scales at which the transition to the zero-temperature
limit, discussed here, occurs (as was done for disordered systems\cite{S1}).
Our results will then be discussed vis-a-vis previous works\cite{vO},\cite
{RUJ}.

\newpage

\section{Figure captions}

\subsubsection{Figure 1}

$\left| \widehat{L}_{if}\right| ^{2}$ vs. ${\cal N}$ ($\varepsilon ={\cal N}$
in our approximation). The number of lines, where the value of $\left| 
\widehat{L}_{if}\right| ^{2}$ might fall, increases with ${\cal N}$. For a
given ${\cal N}$ (more precisely, for a given level $i$), $\left| \widehat{L}
_{if}\right| ^{2}$ is on one of the lines, depending on $\alpha $. As, for
any given $\alpha $, ${\cal N}$ is increased $\left( i\rightarrow i+1=f\text{%
, }f\rightarrow f+1\right) $, $\left| \widehat{L}_{if}\right| ^{2}$
''jumps'' to another line. The lines are the same regardless of $\alpha $.
For illustration, we present numerical results for $\alpha
=8/31^{0.6}\approx 1.01923$. Equations for the three straight lines shown
here are given by (\ref{Lif_straight}).

\subsubsection{Figure 2}

Numerical average of $\left\langle \left| \widehat{L}_{if}\right|
^{2}\right\rangle $ vs. ${\cal N}$ vis-a-vis the fits (\ref{Lif_h}) and (\ref
{Lif_l}). A combined average over $400$ algebraic values of $\alpha \in
\left[ 1,2\right] $ and over the energy interval $\left[ {\cal N}-2\times
10^{4},{\cal N}+2\times 10^{4}\right] $ was used. The bottom line and fit
correspond to the contribution from the top line only in Fig. 1 ($%
k_{x}-n_{x}=\pm 1,k_{y}-n_{y}=\mp 1$).

\subsubsection{Figure 3}

Same as the top set of dots in Fig 2 vs. the result of averaging over a
narrower interval, $\left[ {\cal N}-2\times 10^{2},{\cal N}+2\times
10^{2}\right] $.

\subsubsection{Figure 4}

Shaded areas represent the probabilities $p\left( n_{x}\right) dn_{x}$.

\subsubsection{Figure 5}

Distribution function $p\left( n_{x}\right) $ vs. $n_{x}$: analytical result
with $p\left( n_{x}\right) $ given by eq. (\ref{p_nx}) vis-a-vis numerical
data for $0.99\times 10^{5}\leq \varepsilon \leq 1.01\times 10^{5}$ and the
same $\alpha $ as in Fig. 1.

\subsubsection{Figure 6}

$p\left( k_{x}-n_{x}\right) $ vs. $\left( k_{x}-n_{x}\right) $: analytical
result obtained via (\ref{p_kxnx}) vis-a-vis numerical data for $0.99\times
10^{6}\leq \varepsilon \leq 1.01\times 10^{6}$ for the same $\alpha $ as in
Fig. 1 and for an ensemble average over $100$ algebraic $\alpha \in \left[
1,1.25\right] $.

\subsubsection{Figure 7a}

Probability $P\left( k_{x}-n_{x}=\pm 1\right) $ vs. ${\cal N}$: analytical
ansatz based on (\ref{Pp}) and (\ref{p_0}) and on its approximation (\ref
{p_0-expand}) (solid lines) vis-a-vis numerical data.

\subsubsection{Figure 7b}

$P\left( k_{x}-n_{x}=\pm 1\right) \sqrt{{\cal N}}$ vs. ${\cal N}$: same as
Fig.7a, but with the main dependence, $\propto {\cal N}^{-1/2}$, eliminated.

\subsubsection{Figure 8}

Distribution functions of conditional probability $p\left( k_{y}-n_{y}\mid
k_{x}-n_{x}=-1\right) $ and $p\left( \left| k_{y}-n_{y}\right| \mid
k_{x}-n_{x}=1\right) $: analytical result obtained via (\ref{p_kyny}) vs.
numerical data. Notice that for the latter $p\left( 0\mid
k_{x}-n_{x}=-1\right) =0$ while $p\left( 0\mid k_{x}-n_{x}=1\right) \neq 0$,
as explained in text.

\subsubsection{Figure 9}

Conditional probability $P\left( k_{y}-n_{y}=\mp 1\mid k_{x}-n_{x}=\pm
1\right) $ vs. ${\cal N}$: an ansatz based on $0.225\times p\left( 0\mid
\left| k_{x}-n_{x}\right| =1\right) $ and its approximation $0.14\times
\left( 1+\Delta N/4N\right) $ (solid lines) vis-a-vis numerical data.

\subsubsection{Figure 10}

Total magnetic susceptibility, evaluated via (\ref{khi_tot}) and (\ref{diam}%
) - top set of dots. $\left\langle 2\left| \widehat{L}_{if}\right|
^{2}\left( \varepsilon _{f}-\varepsilon _{i}\right) ^{-1}\right\rangle $,
where $\varepsilon _{i}$ and $\varepsilon _{f}$ are the Fermi level and the
first unoccupied level respectively - middle set of dots. $\left\langle
2\left| \widehat{L}_{if}\right| ^{2}\left( \varepsilon _{f}-\varepsilon
_{i}\right) ^{-1}\right\rangle _{top}$, where $\varepsilon _{i}$ and $%
\varepsilon _{f}$ are the Fermi level and the first unoccupied level such
that $k_{x}-n_{x}=\pm 1,k_{y}-n_{y}=\mp 1$ (top line in Fig. 1) - bottom set
of dots. $2\left\langle \left| \widehat{L}_{if}\right| ^{2}\right\rangle
\left\langle \left( \varepsilon _{f}-\varepsilon _{i}\right)
^{-1}\right\rangle $ with $\left\langle \left| \widehat{L}_{if}\right|
^{2}\right\rangle $ given by the top set of dots in Fig. 2 and $\left\langle
\left( \varepsilon _{f}-\varepsilon _{i}\right) ^{-1}\right\rangle \approx
15.5$. A combined average over $400$ algebraic values of $\alpha \in \left[
1,2\right] $ and over the energy interval $\left[ {\cal N}-2\times 10^{3},%
{\cal N}+2\times 10^{3}\right] $ was used.

\subsubsection{Figure 11}

Large dots same as top dots in Fig. 10. Small dots - a combined average over
same $400$ algebraic values of $\alpha \in \left[ 1,2\right] $ but over the
narrower energy interval $\left[ {\cal N}-2\times 10^{2},{\cal N}+2\times
10^{2}\right] $.

\subsubsection{Figure 12a}

$\left( \varepsilon _{f}-\varepsilon _{i}\right) ^{-1}$ vs. ${\cal N}$ for $%
\alpha _{1}=457^{0.855}/185\approx 1.01638$, where $\varepsilon _{i}$ and $%
\varepsilon _{f}$ are the nearest levels such that $k_{x}-n_{x}=\pm
1,k_{y}-n_{y}=\mp 1$. $\beta =$ $\alpha -1$ is decomposed according to the
first eq. (\ref{beta_examples}). Coordinates of point $A$ are $\left(
718964,25443.6\right) $ and can be described analytically by (\ref
{ei_ef_top_rat2}) with $t=-62$, $C=n_{x}-n_{y}=10$ and $n_{y}=672$.

\subsubsection{Figure 12b}

Same as Figure 12a, shows $\varepsilon _{f}-\varepsilon _{i}$ vs. ${\cal N}$
and point $A$.

\subsubsection{Figure 13a}

$\left( \varepsilon _{f}-\varepsilon _{i}\right) ^{-1}$ vs. ${\cal N}$ for $%
\alpha _{2}=911^{0.755}/150\approx 1.14379$ and $\beta $ given by second eq.
(\ref{ei_ef_top_rat2}) Points $A$-$D$ have been examined; they can be
described analytically by equations that are an extension of (\ref
{ei_ef_top_rat2}) to include the next iteration in (\ref{beta_truncated}).

\subsubsection{Figure 13b}

Same as Figure 13a, shows $\varepsilon _{f}-\varepsilon _{i}$ vs. ${\cal N}$
and points $A$-$D$.

\subsubsection{Figure 14a}

$\left( \varepsilon _{f}-\varepsilon _{i}\right) ^{-1}$ vs. ${\cal N}$ for $%
\alpha _{3}=643^{0.655}/63\approx 1.09657$ and $\beta $ given by third eq. (%
\ref{ei_ef_top_rat2}). Points $A$ and $B$ have been examined; they can be
described analytically by equations that are an extension of (\ref
{ei_ef_top_rat2}) to include the next iteration in (\ref{beta_truncated}).

\subsubsection{Figure 14b}

Same as Figure 14a, shows $\varepsilon _{f}-\varepsilon _{i}$ vs. ${\cal N}$
and points $A$ and $B$.

\subsubsection{Figure 15a}

$\left( \varepsilon _{f}-\varepsilon _{i}\right) ^{-1}$ vs. ${\cal N}$ for
the rational aspect ratio $\alpha =1.01$. All the points are described
analytically by eqs. (\ref{ei_ef_top_rat1}) with $p=100$. The maxima,
denoted by $A$ and $B$ (two points each, corresponding to increasing and
decreasing functions of ${\cal N}$ in (\ref{ei_ef_top_rat1})) are given by
eqs. (\ref{ei_ef_top_rat1-min}).

\subsubsection{Figure 15b}

Same as Figure 15a, shows $\varepsilon _{f}-\varepsilon _{i}$ vs. ${\cal N}$
and points $A$ and $B$.

\resizebox{5in}{!}{\rotatebox{90}{\includegraphics{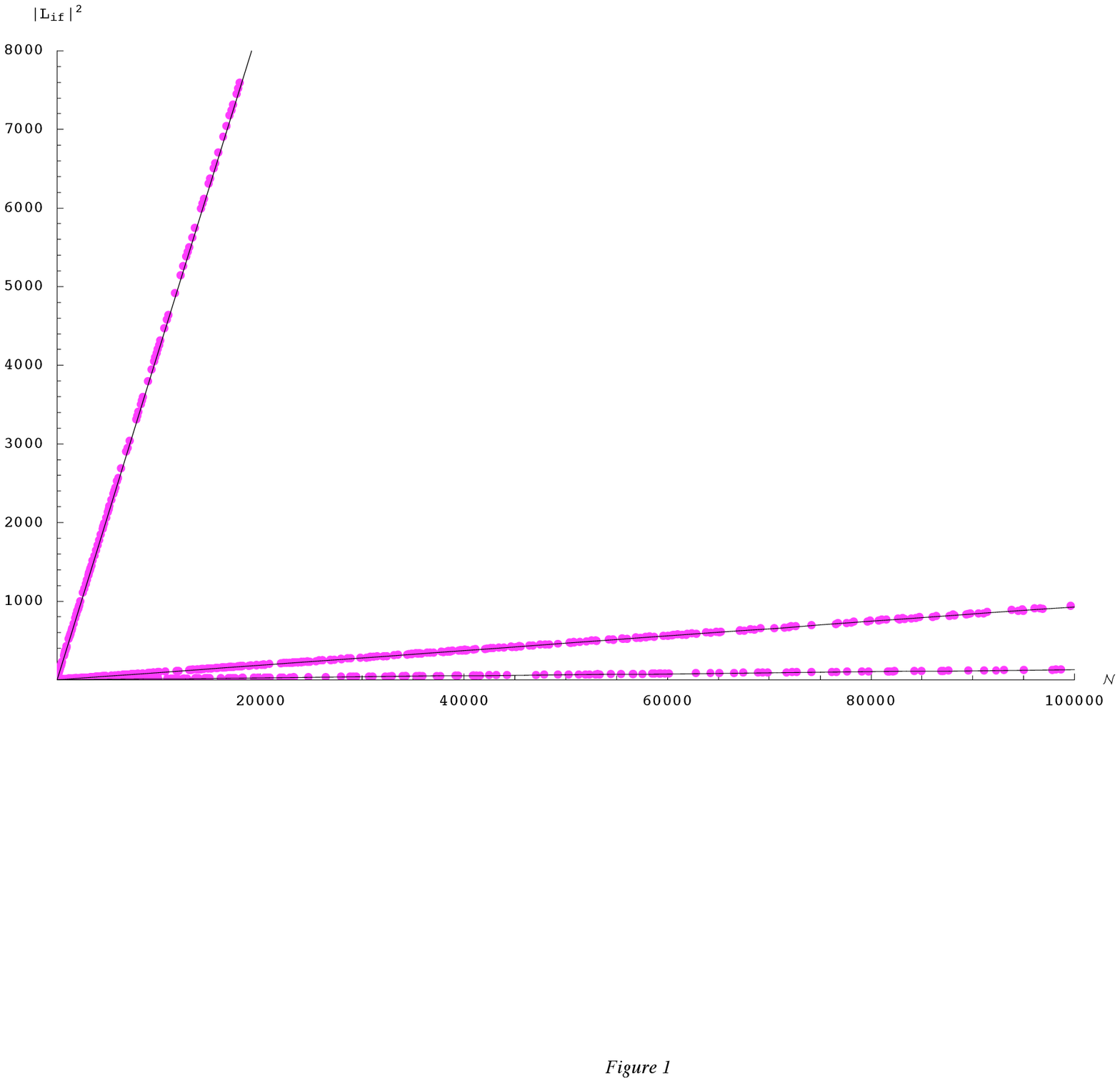}}}
\clearpage \resizebox{5in}{!}{\rotatebox{90}{\includegraphics{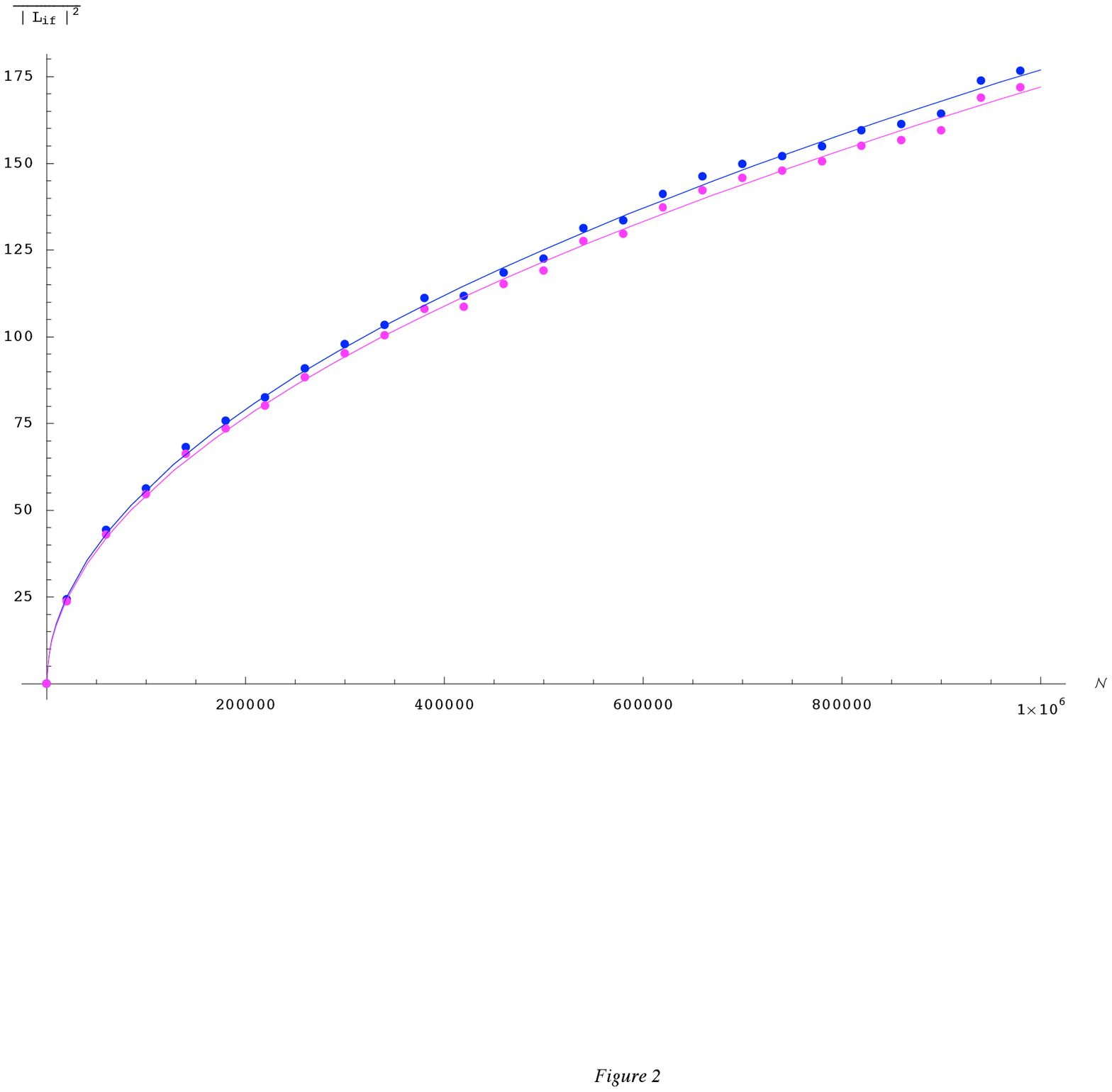}}}
\clearpage \resizebox{5in}{!}{\rotatebox{90}{\includegraphics{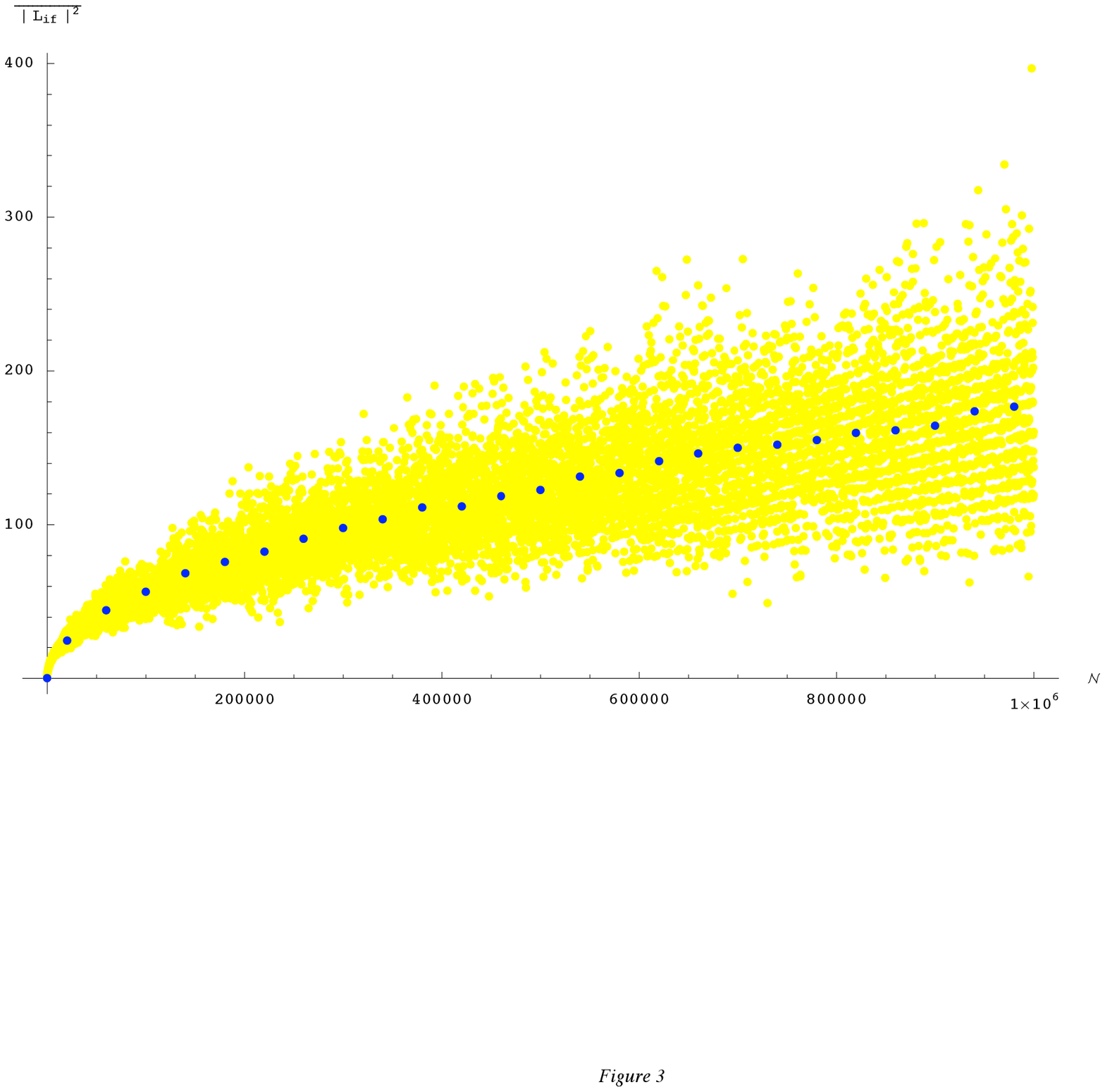}}}
\clearpage \resizebox{5in}{!}{\rotatebox{90}{\includegraphics{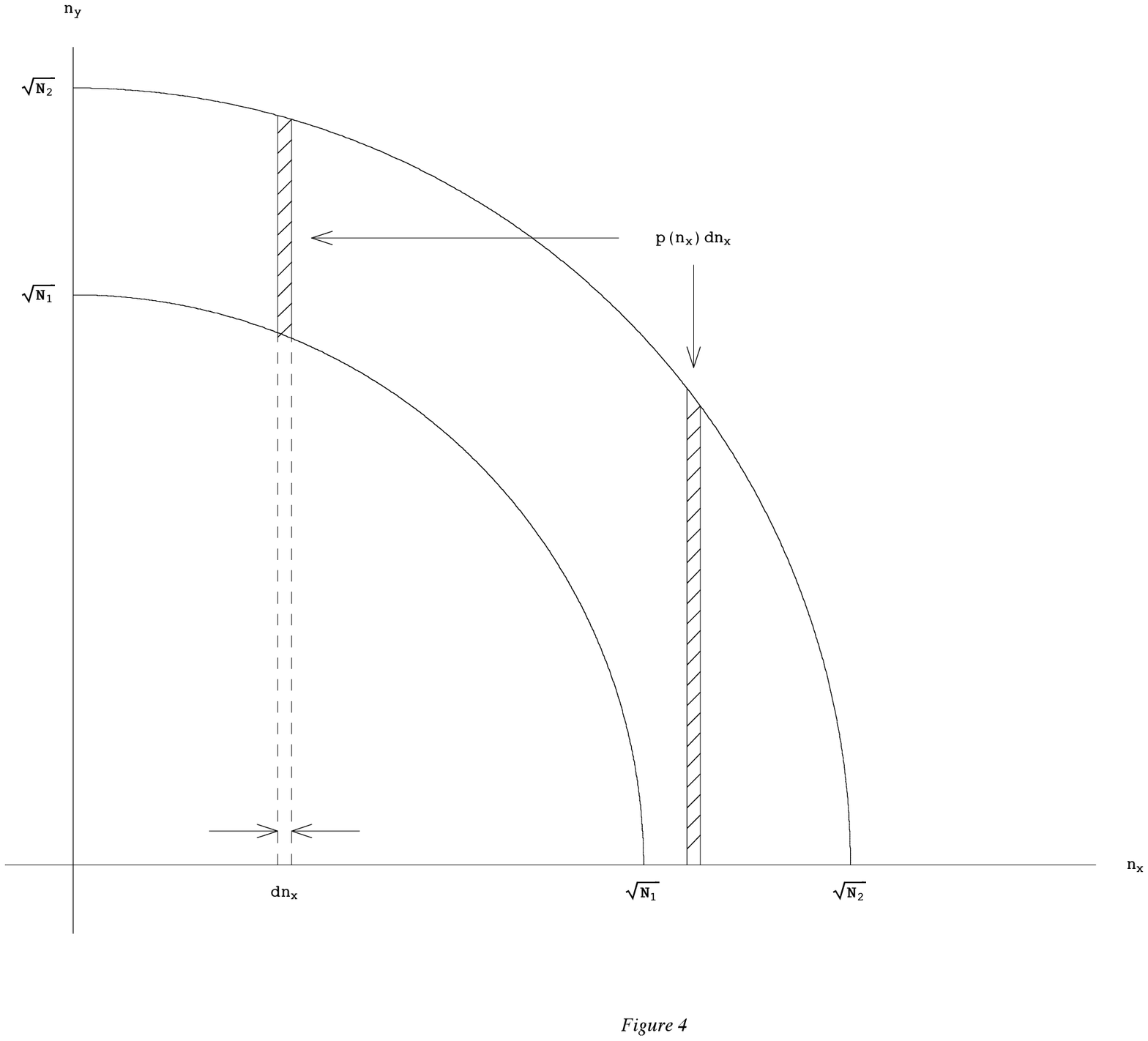}}}
\clearpage \resizebox{5in}{!}{\rotatebox{90}{\includegraphics{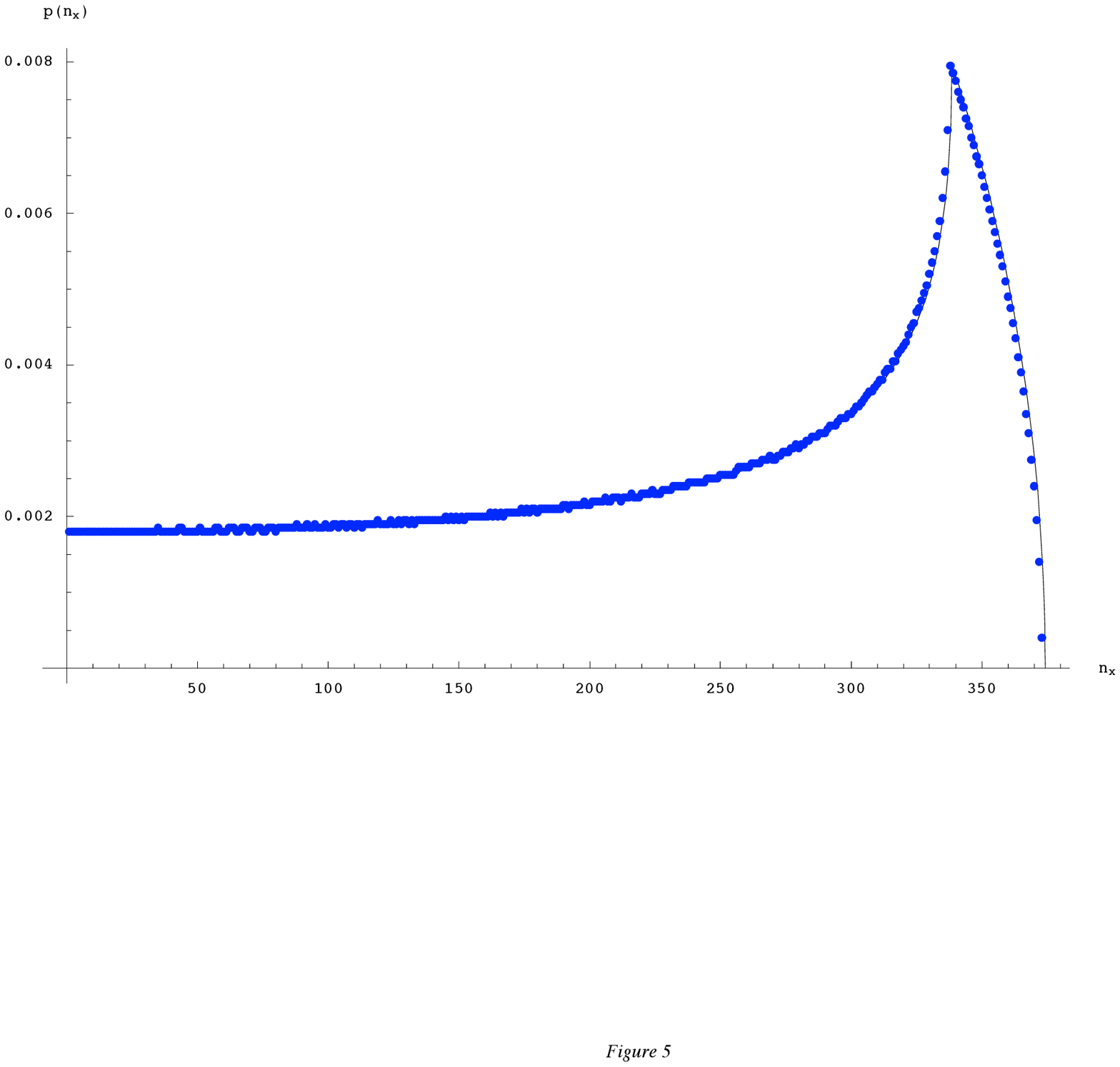}}}
\clearpage \resizebox{5in}{!}{\rotatebox{90}{\includegraphics{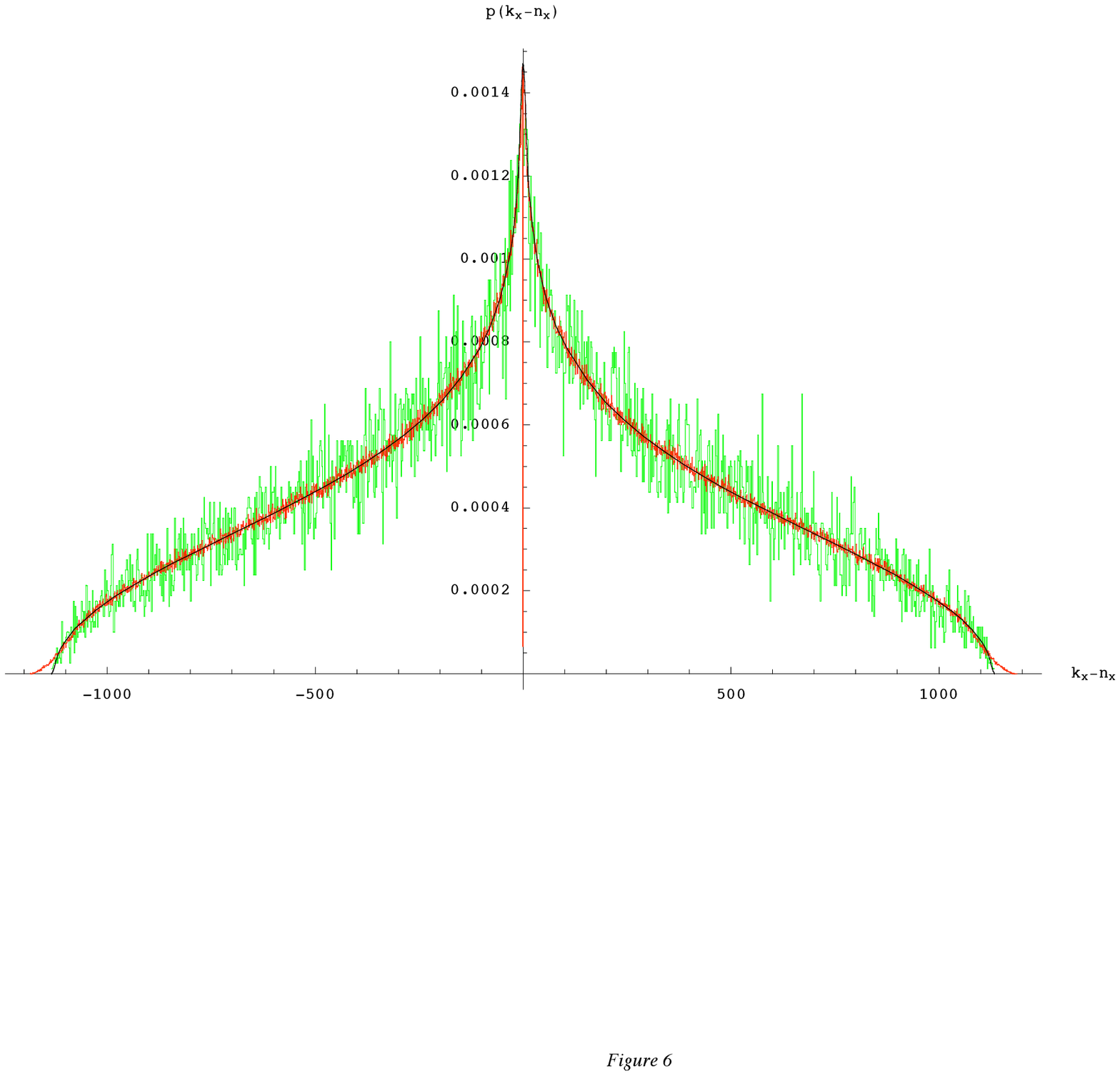}}}
\clearpage \resizebox{5in}{!}{\rotatebox{90}{\includegraphics{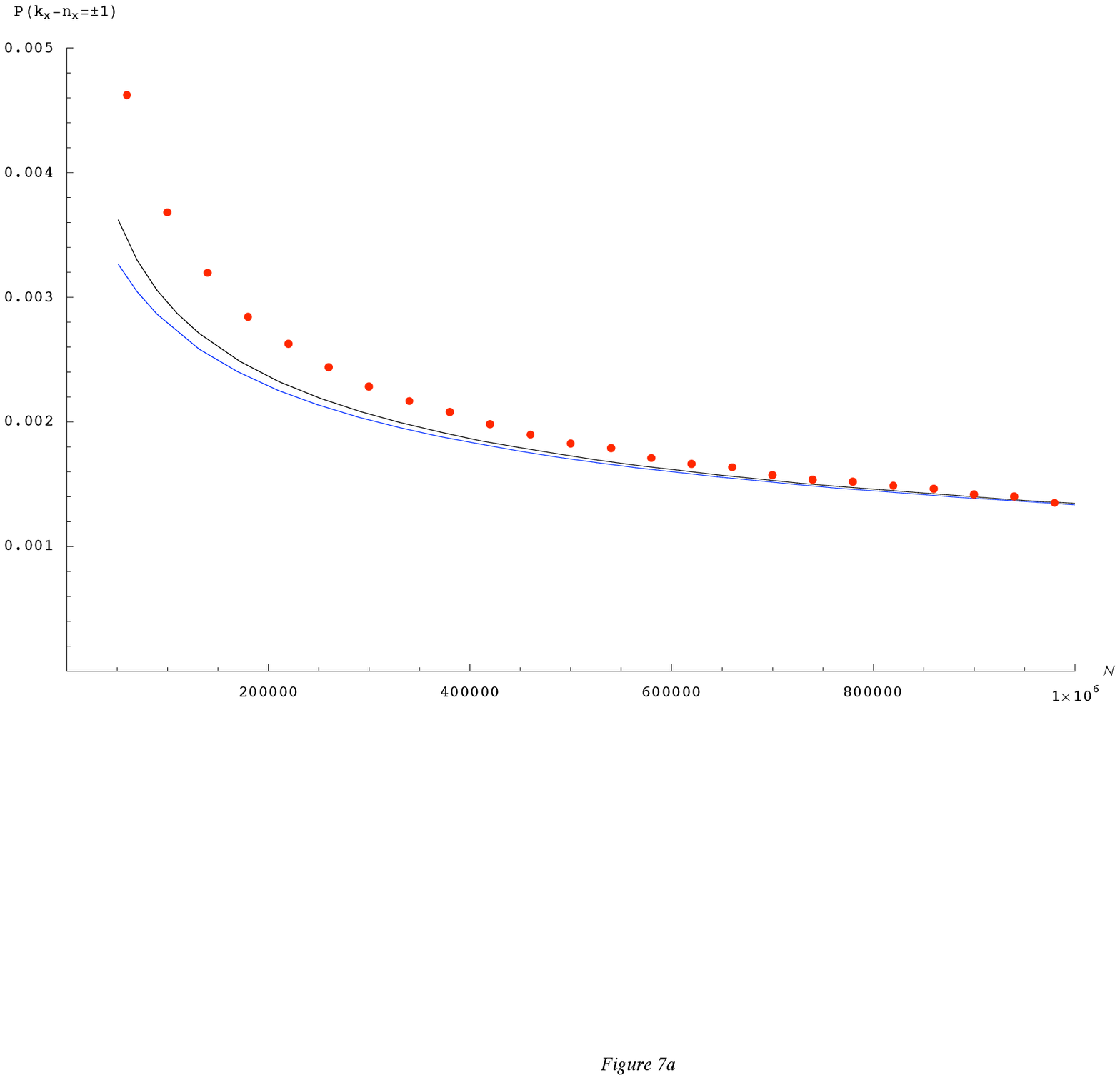}}}
\clearpage \resizebox{5in}{!}{\rotatebox{90}{\includegraphics{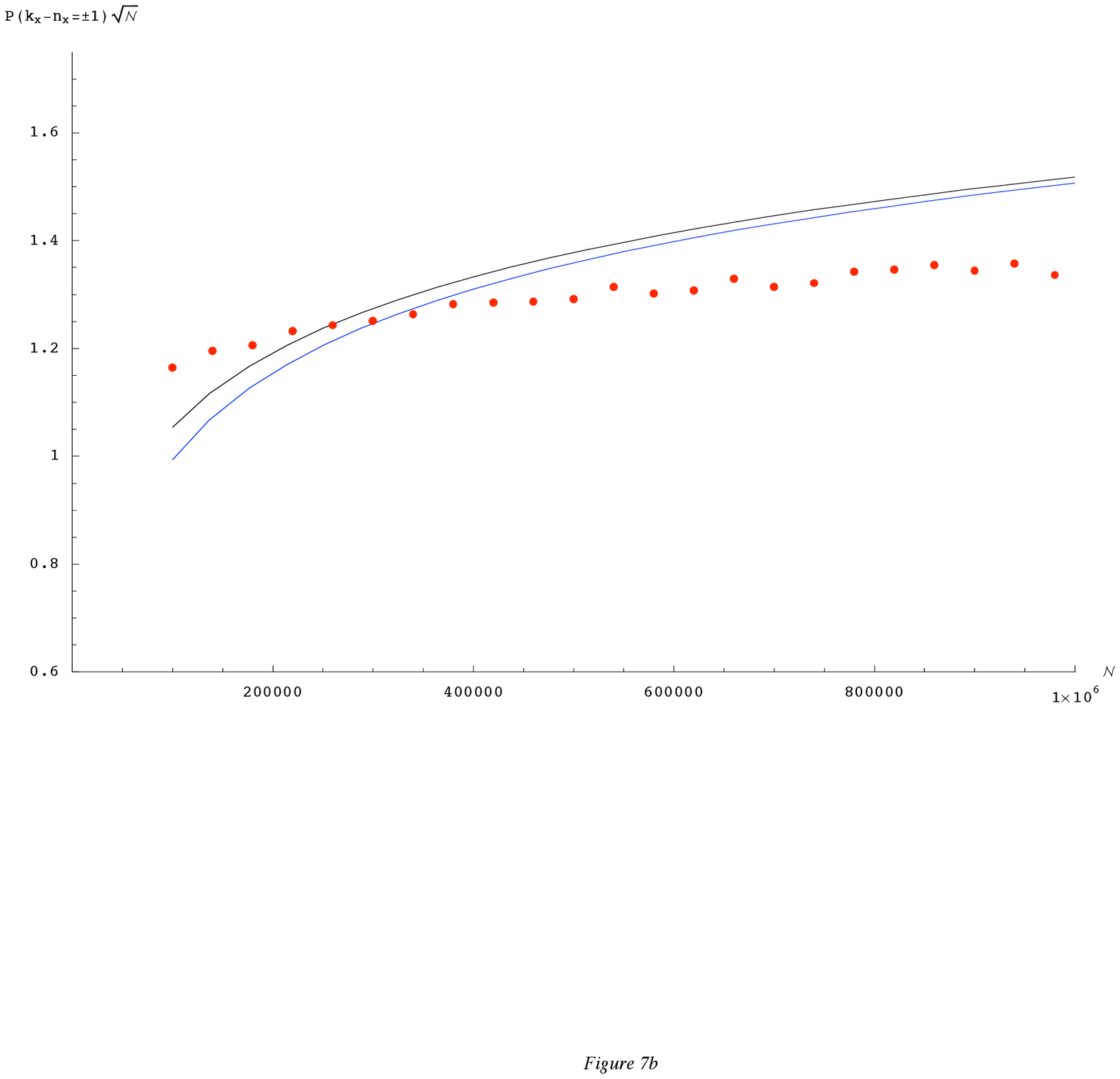}}}
\clearpage \resizebox{5in}{!}{\rotatebox{90}{\includegraphics{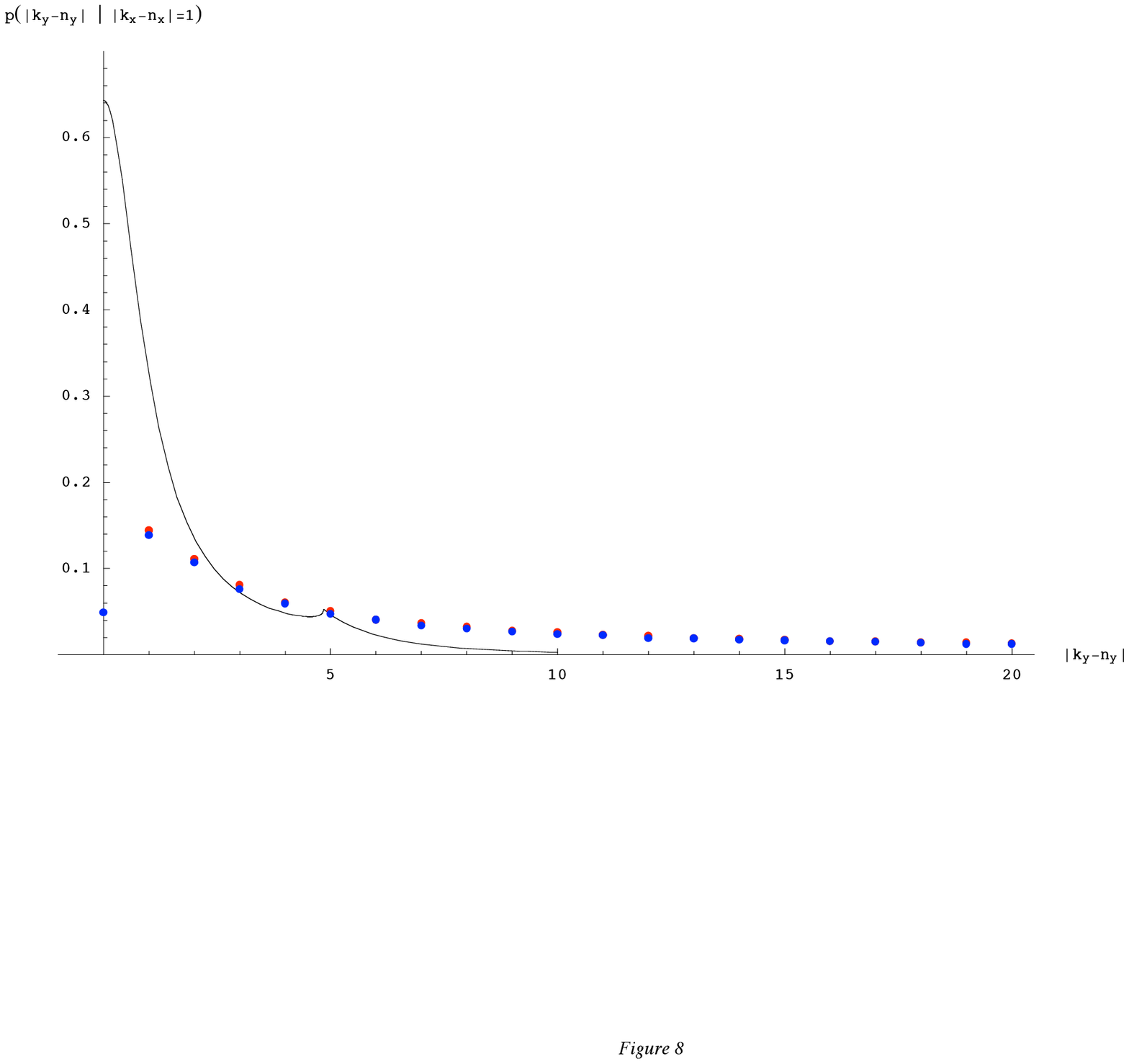}}}
\clearpage \resizebox{5in}{!}{\rotatebox{90}{\includegraphics{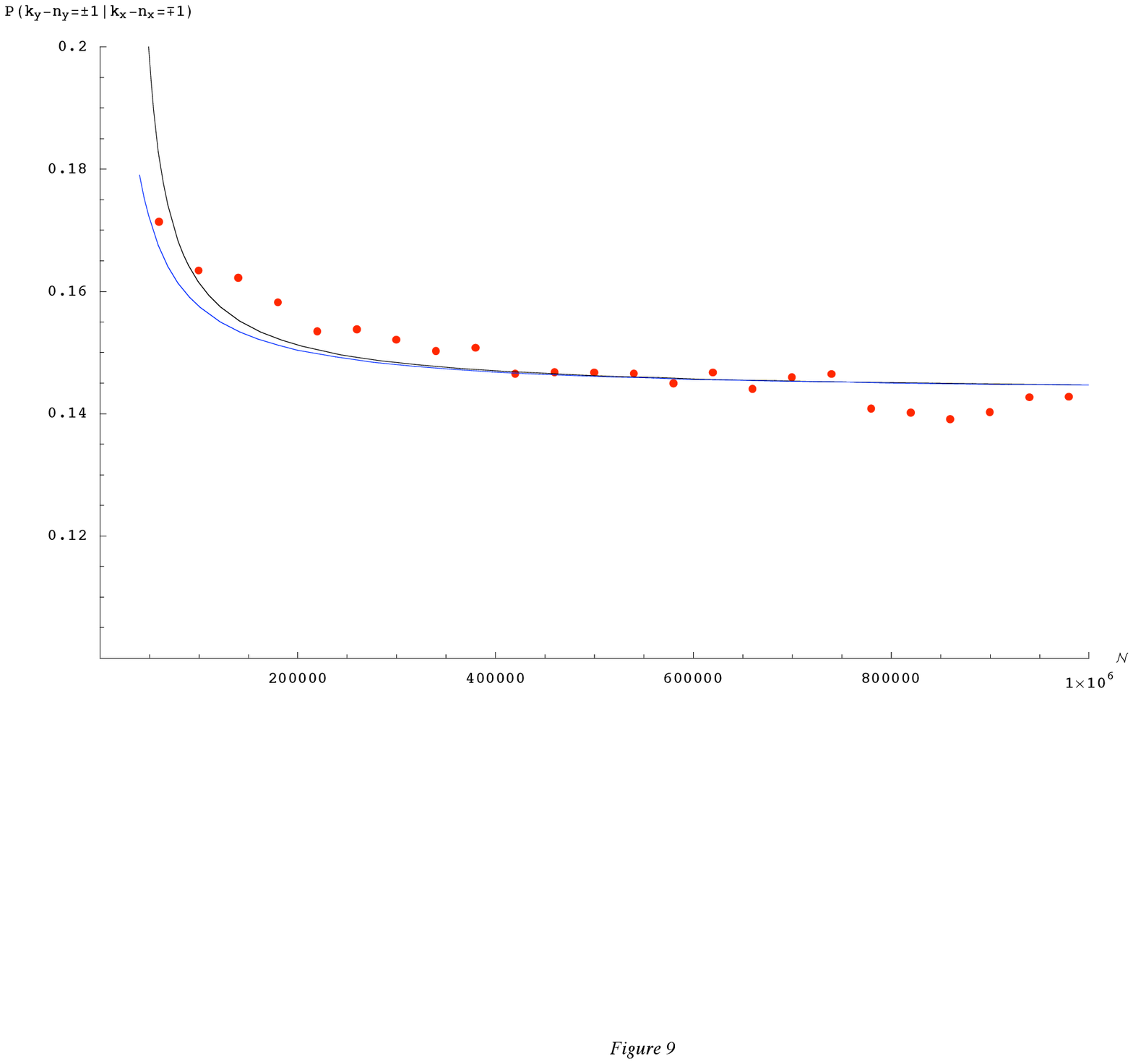}}}
\clearpage \resizebox{5in}{!}{\rotatebox{90}{\includegraphics{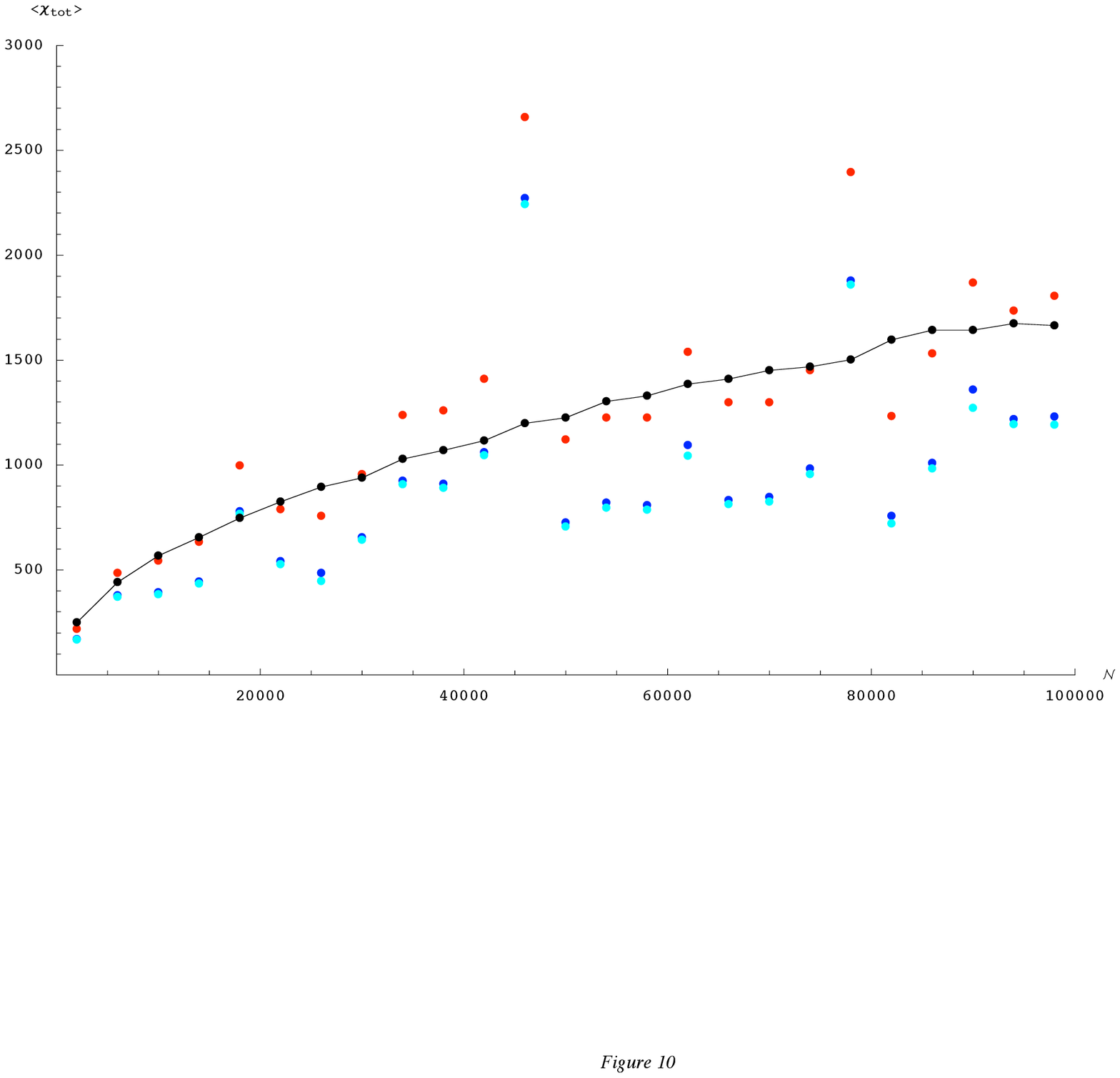}}}
\clearpage \resizebox{5in}{!}{\rotatebox{90}{\includegraphics{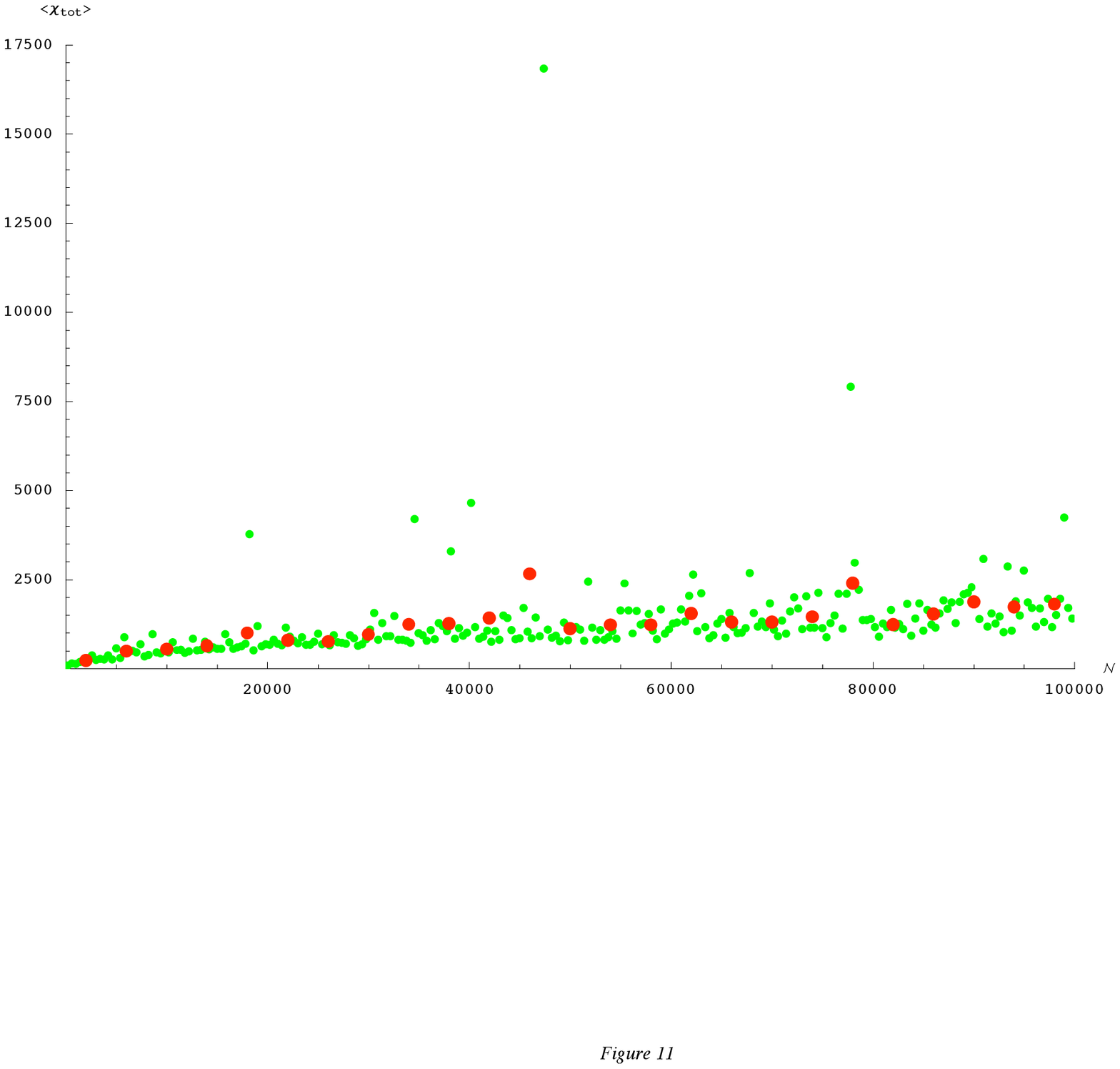}}}
\clearpage \resizebox{5in}{!}{\rotatebox{90}{\includegraphics{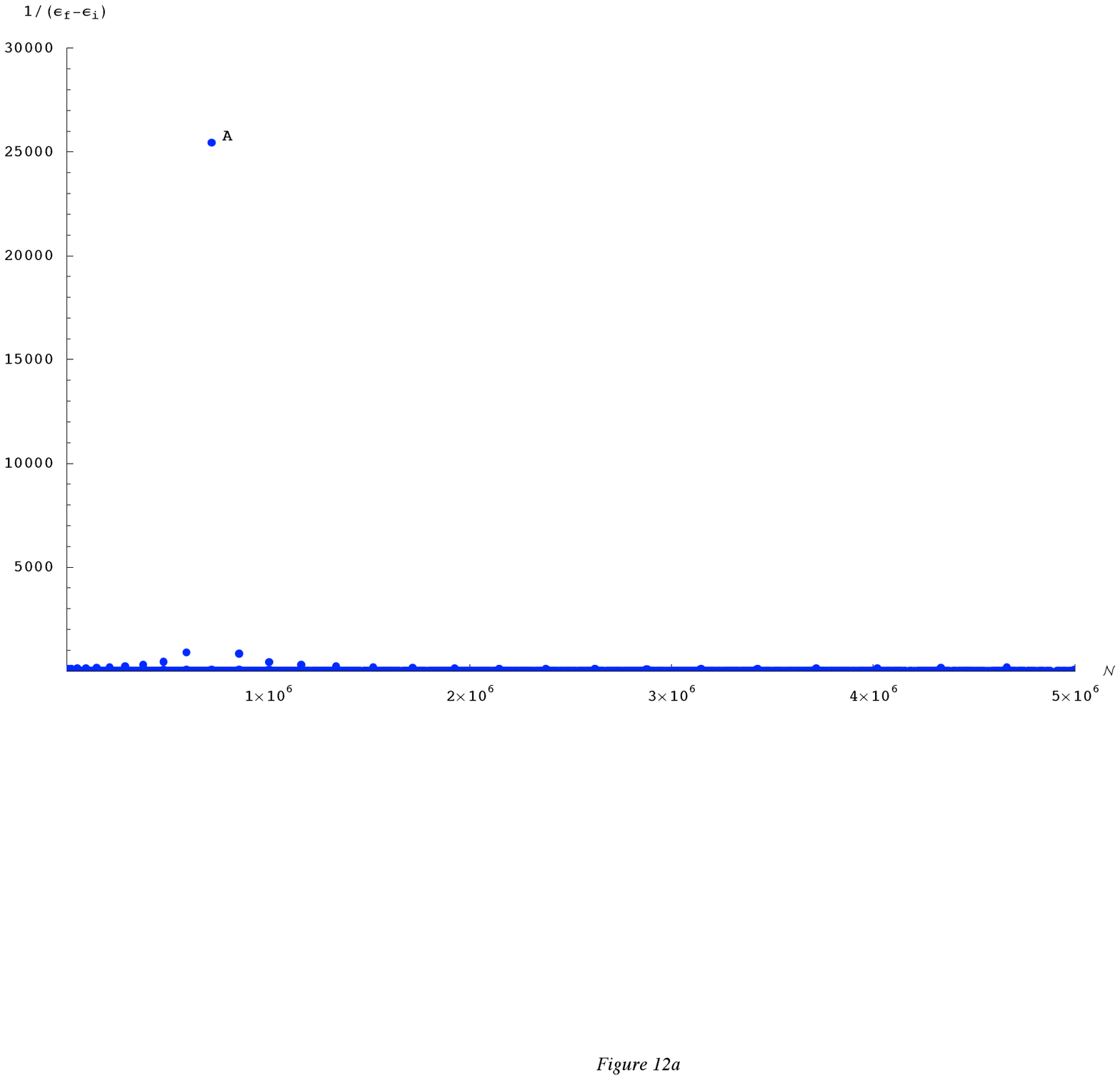}}}
\clearpage \resizebox{5in}{!}{\rotatebox{90}{\includegraphics{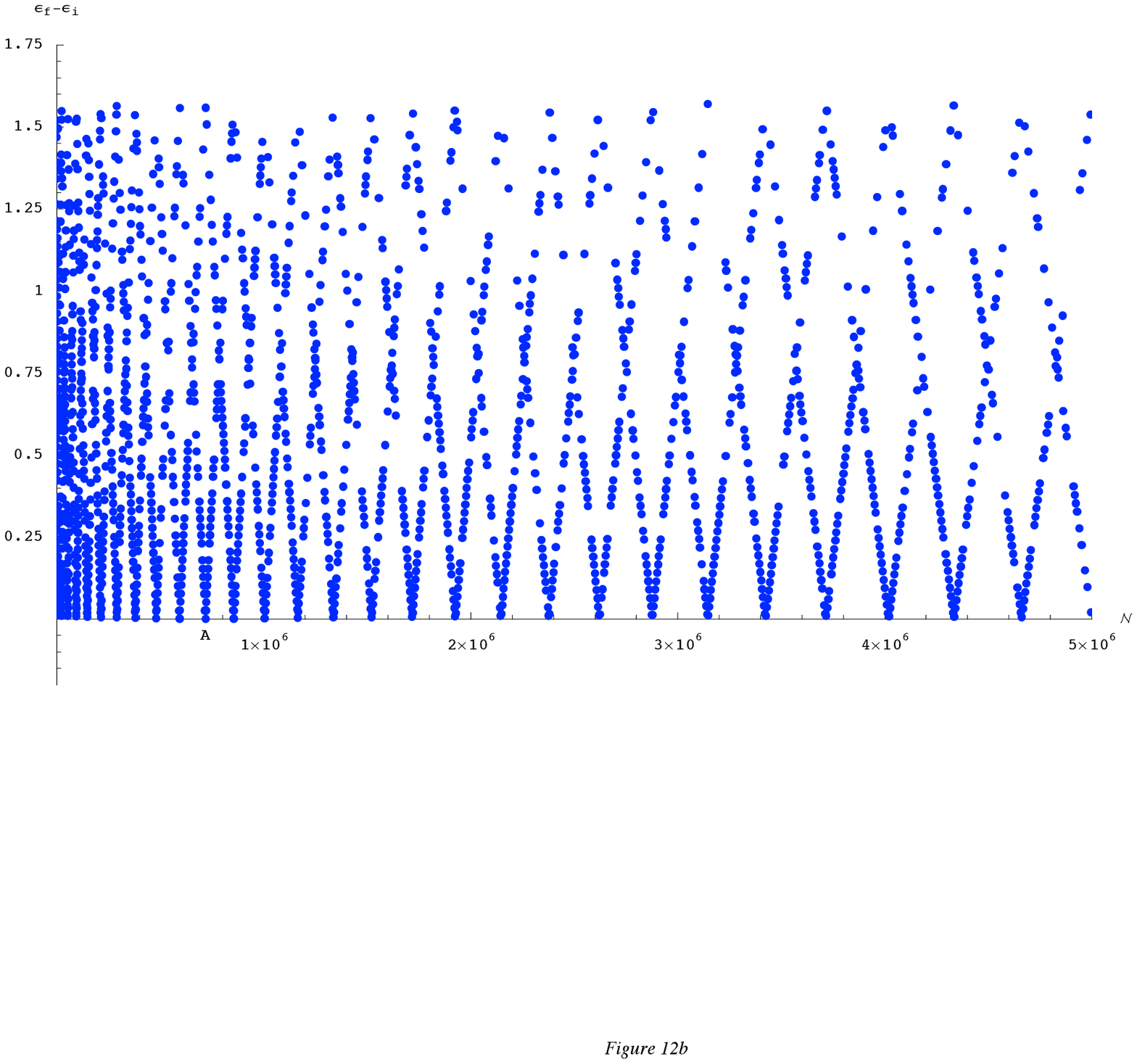}}}
\clearpage \resizebox{5in}{!}{\rotatebox{90}{\includegraphics{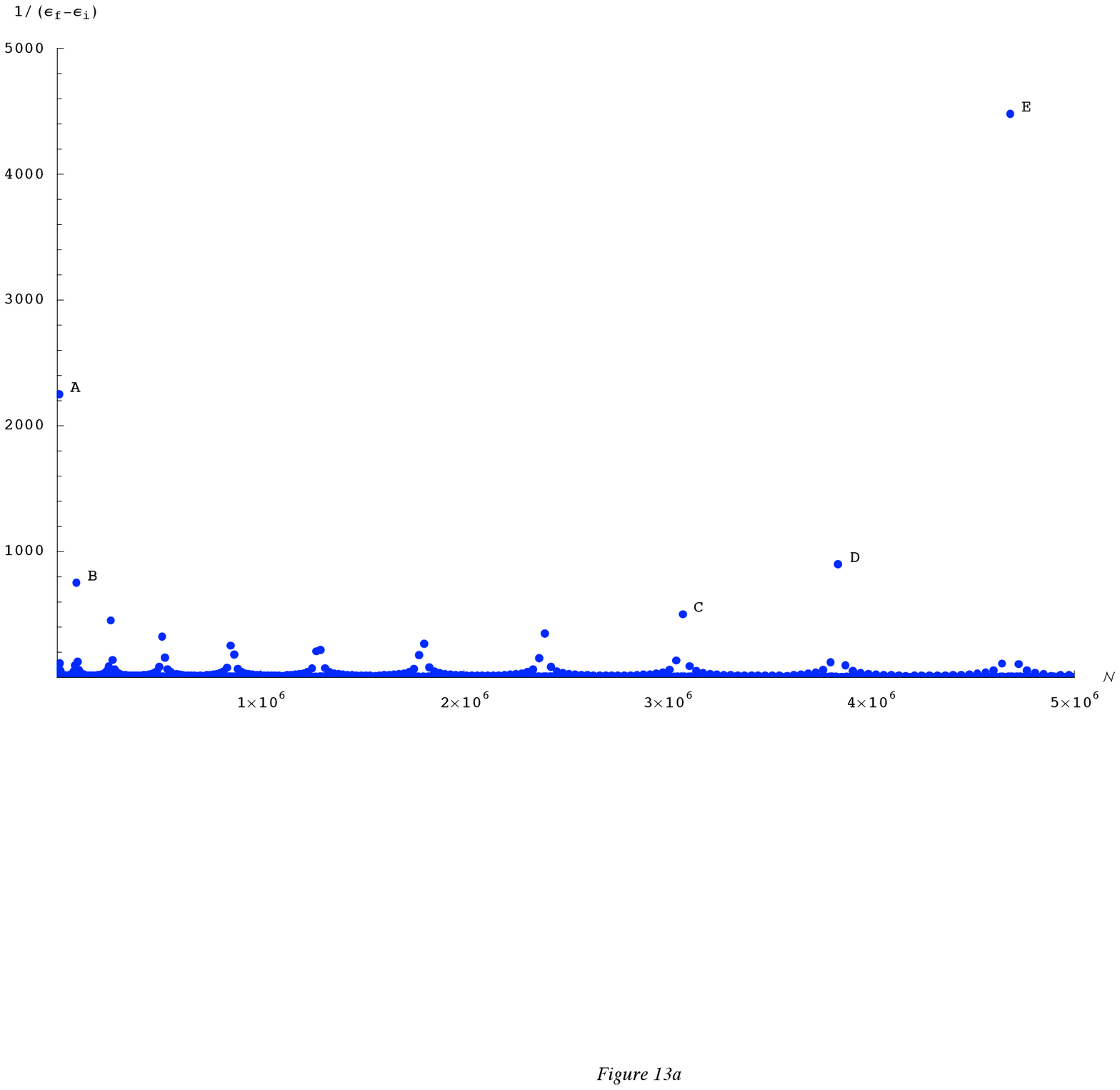}}}
\clearpage \resizebox{5in}{!}{\rotatebox{90}{\includegraphics{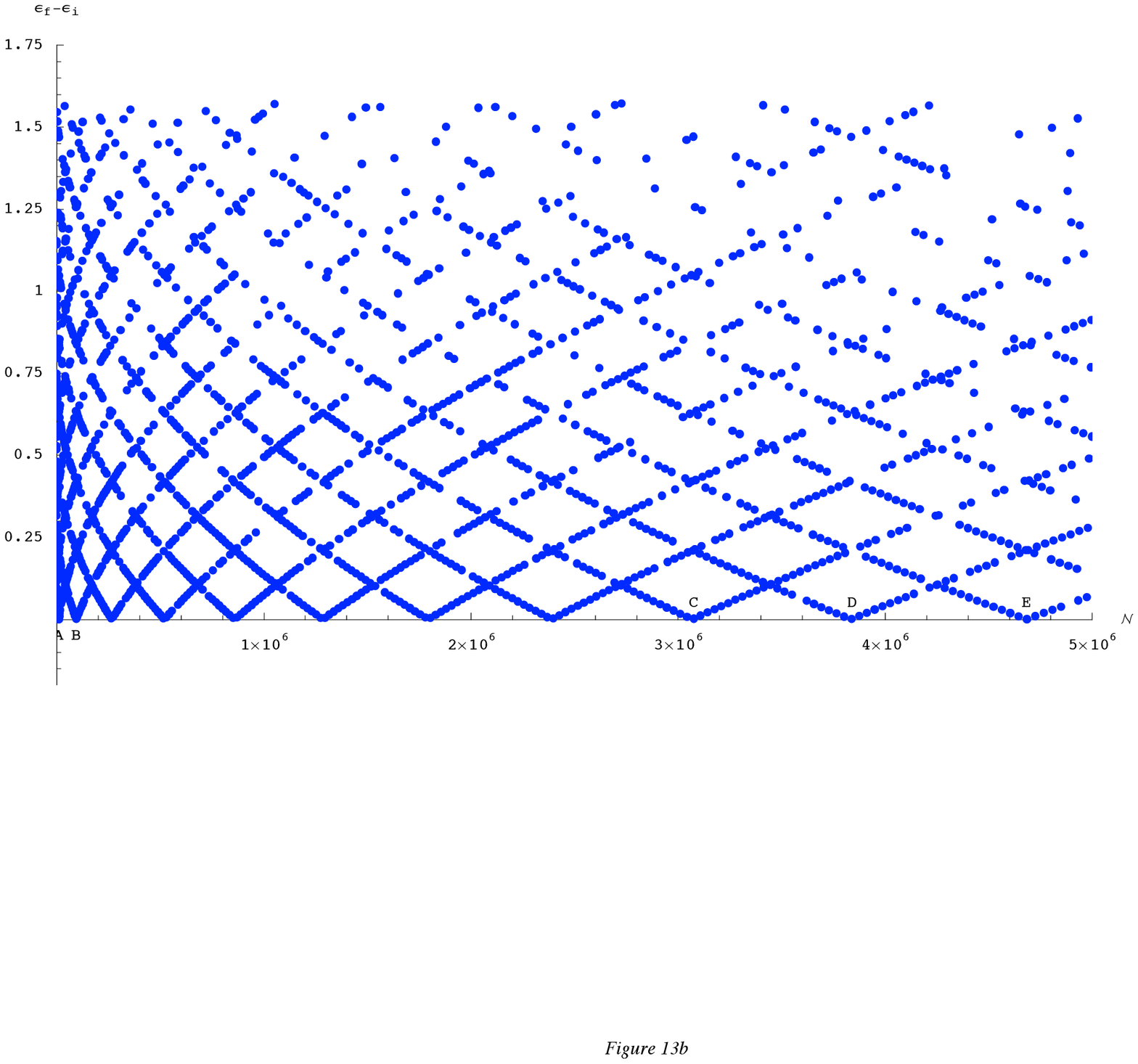}}}
\clearpage \resizebox{5in}{!}{\rotatebox{90}{\includegraphics{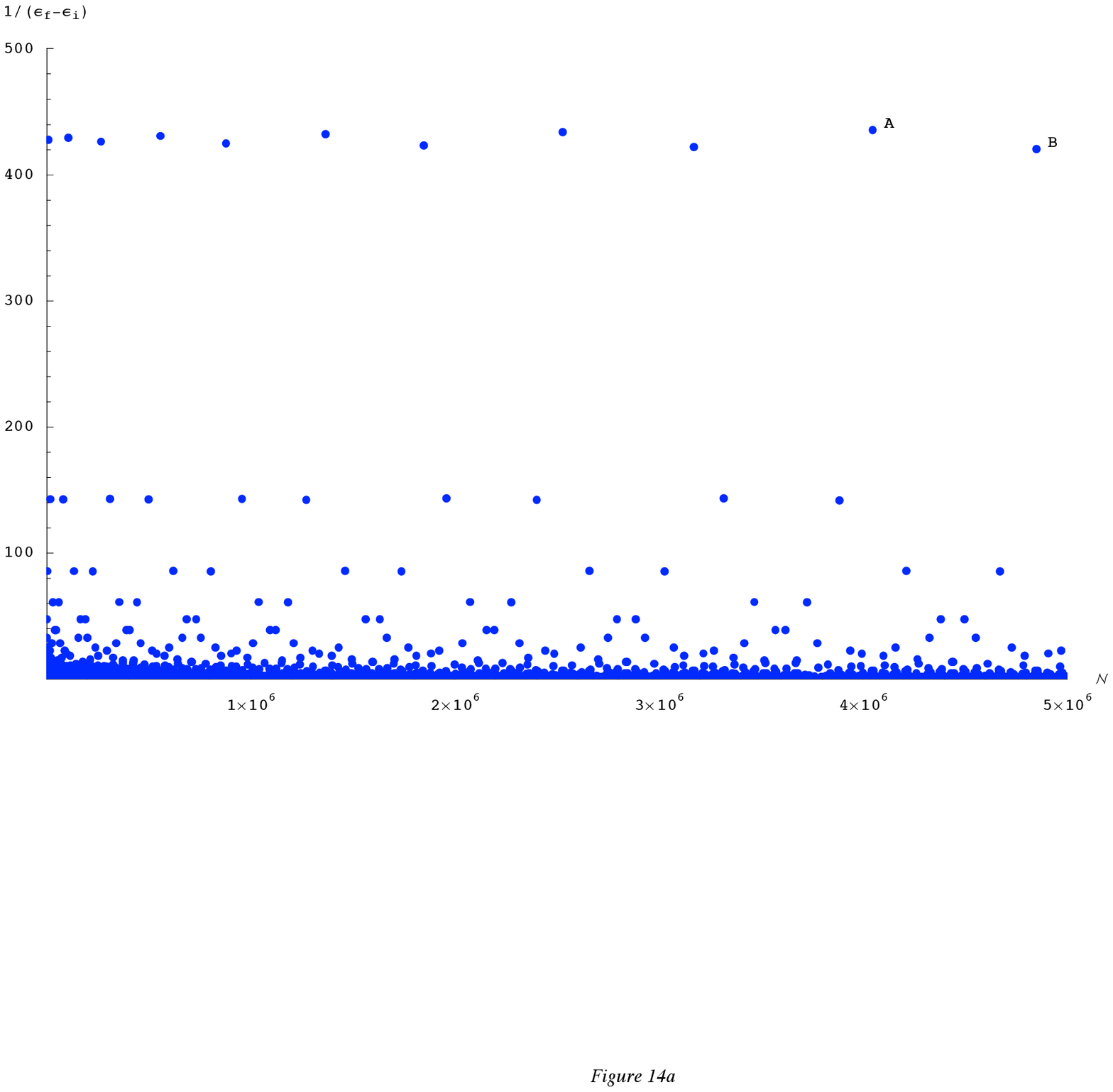}}}
\clearpage \resizebox{5in}{!}{\rotatebox{90}{\includegraphics{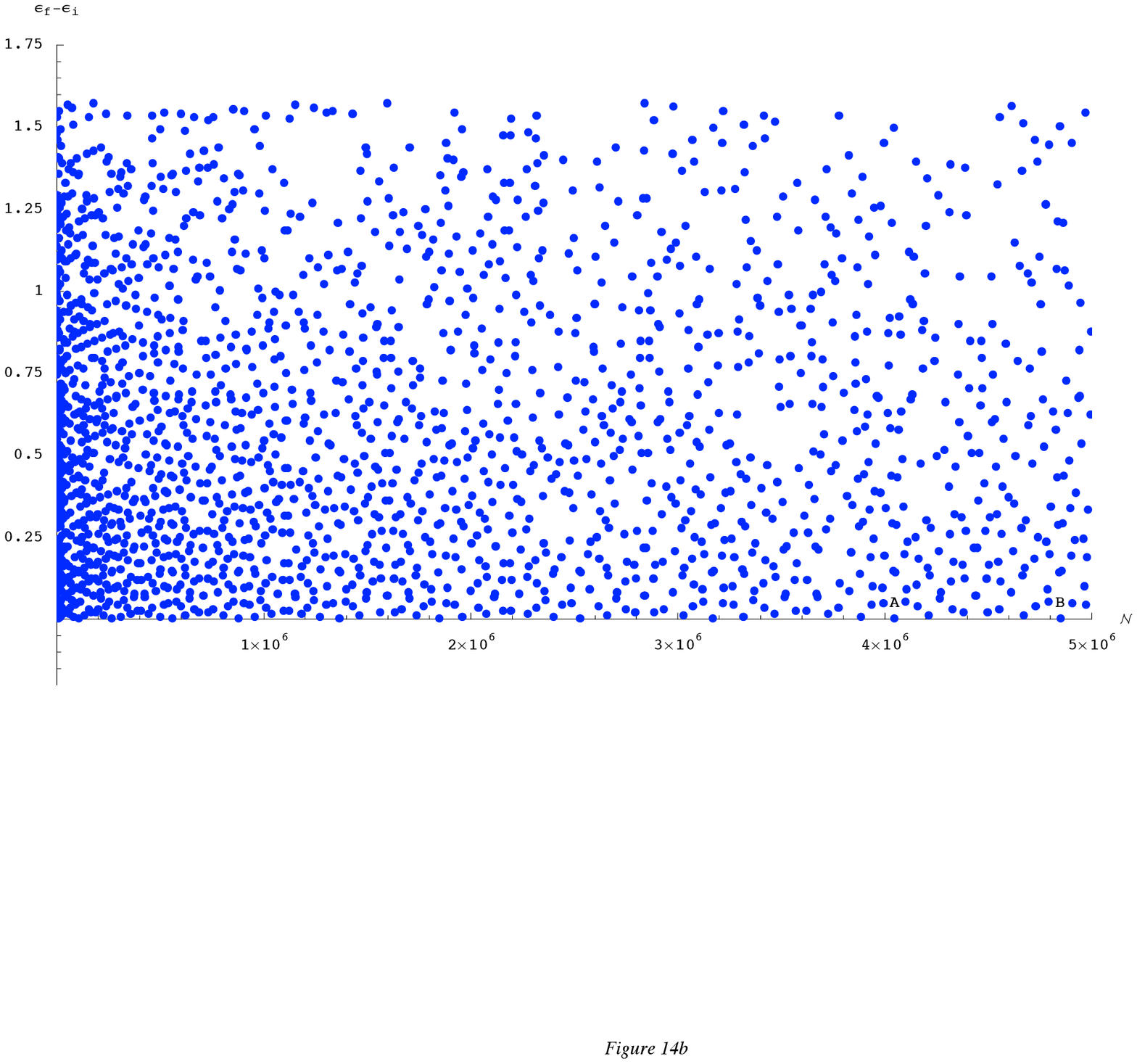}}}
\clearpage \resizebox{5in}{!}{\rotatebox{90}{\includegraphics{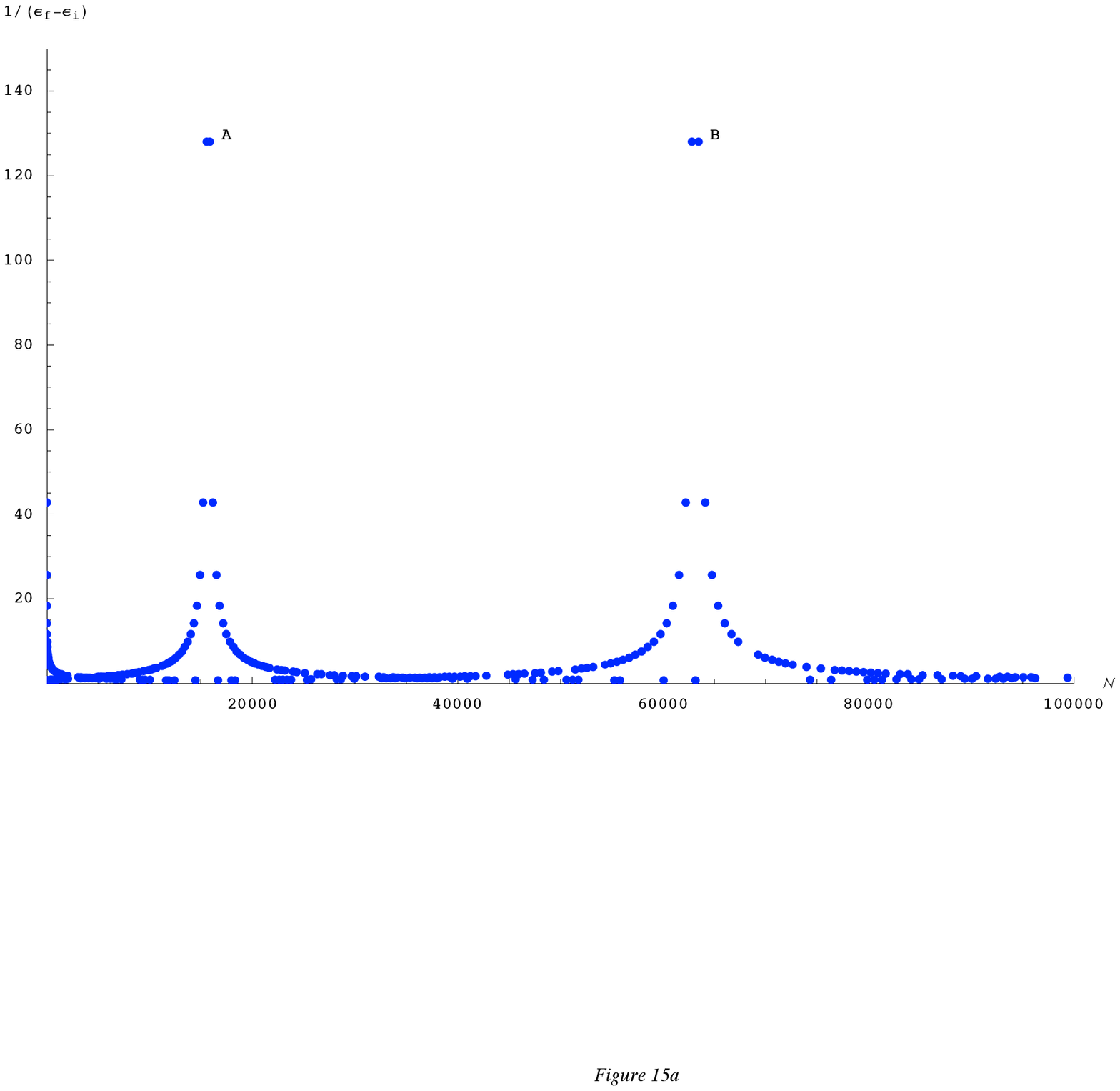}}}
\clearpage \resizebox{5in}{!}{\rotatebox{90}{\includegraphics{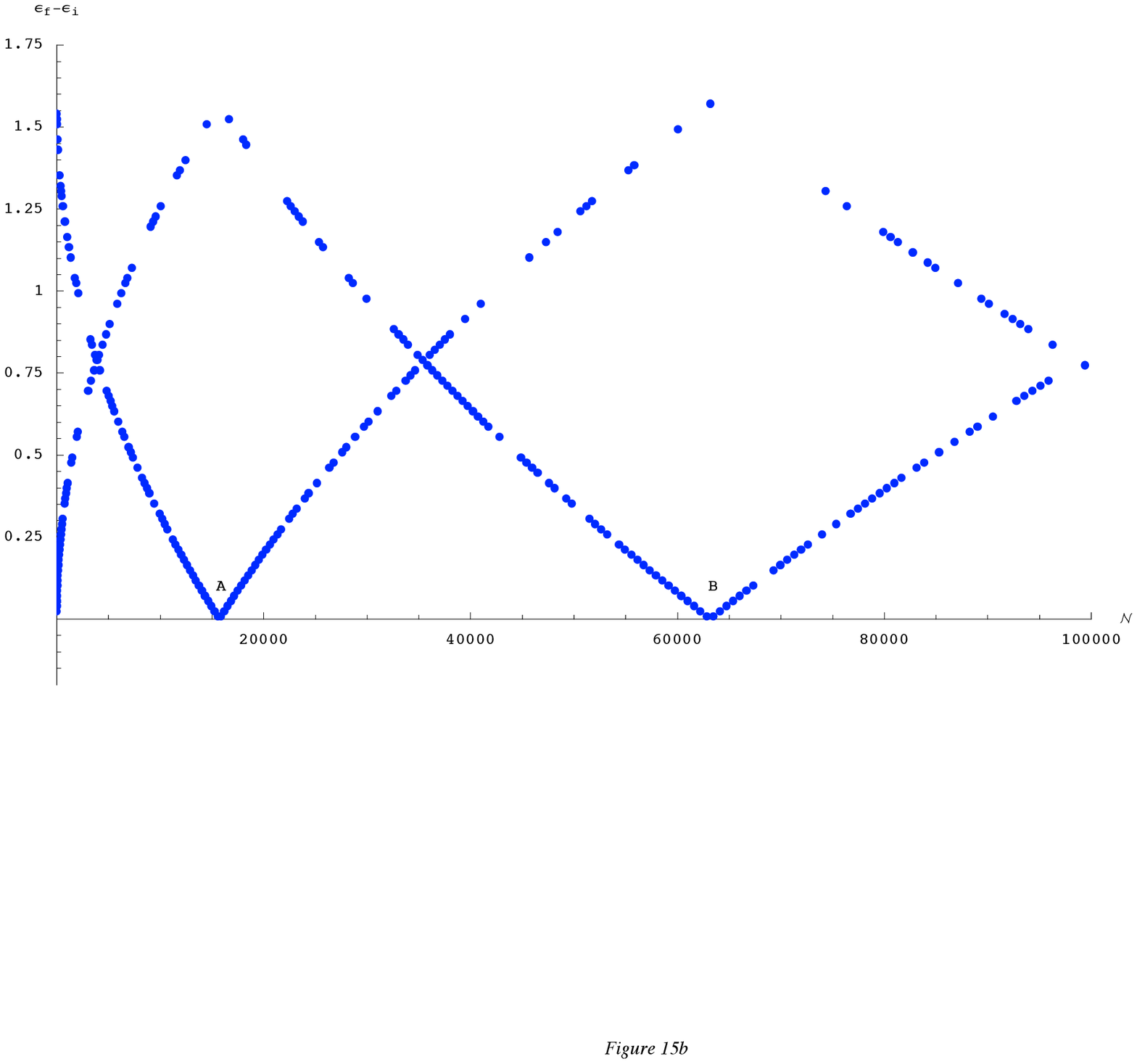}}}

\end{document}